\documentclass[9pt,nofootinbib,superscriptaddress,longbibliography,twocolumn]{revtex4-1}
\usepackage{times}
\usepackage{graphicx}
\usepackage{dcolumn}
\usepackage{bm,bbm}
\usepackage{tabularx}
\usepackage{xcolor}
\usepackage{tcolorbox}
\usepackage{algorithm}
\usepackage{algpseudocode}
\usepackage{amsmath}
\usepackage{bbold}
\usepackage{braket}
\usepackage{nameref}
\usepackage[toc,page]{appendix}

\usepackage[colorlinks,breaklinks,linkcolor={blue},citecolor={magenta},urlcolor={blue}]{hyperref}

\makeatletter
\newcommand\org@hypertarget{}
\let\org@hypertarget\hypertarget
\renewcommand\hypertarget[2]{%
  \Hy@raisedlink{\org@hypertarget{#1}{}}#2%
  }
\makeatother

\begin{document} 

\title{Quantum Face Recognition Protocol with Ghost Imaging}

\author{Vahid Salari}
\email{vsalari@bcamath.org}
\affiliation{BCAM - Basque Center for Applied Mathematics; Alameda de Mazarredo, 14,48009 Bilbao, Basque Country - Spain}
\affiliation{Department of Physics and Astronomy, Howard University, Washington DC, USA}

\author{Dilip Paneru}
\affiliation{Department of physics, University of Ottawa, Advanced Research Complex, 25 Templeton Street, K1N 6N5, Ottawa, ON, Canada}

\author{Erhan Saglamyurek}
\affiliation{Department of Physics, University of Alberta, Edmonton, AB T6G 2E1, Canada}
\affiliation{Department of Physics and Astronomy, University of Calgary, Calgary, AB T2N 1N4 Canada}

\author{Milad Ghadimi}
\affiliation{Department of Physics, Isfahan University of Technology, Isfahan 8415683111, Iran}

\author{Moloud Abdar}
\affiliation{Institute for Intelligent Systems Research and Innovation (IISRI), Deakin University, Australia}

\author{Mohammadreza Rezaee}
\affiliation{Department of physics, University of Ottawa, Advanced Research Complex, 25 Templeton Street, K1N 6N5, Ottawa, ON, Canada}

\author{Mehdi Aslani}
\affiliation{Department of Physics, Isfahan University of Technology, Isfahan 8415683111, Iran}

\author{Shabir Barzanjeh}
\affiliation{Department of Physics and Astronomy, University of Calgary, Calgary, AB T2N 1N4 Canada}

\author{Ebrahim Karimi}
\email{ekarimi@uottawa.ca}
\affiliation{Department of physics, University of Ottawa, Advanced Research Complex, 25 Templeton Street, K1N 6N5, Ottawa, ON, Canada}
\affiliation{National Research Council of Canada, 100 Sussex Drive, K1A 0R6, Ottawa, ON, Canada}

\begin{abstract}
\noindent\textbf{Face recognition is one of the most ubiquitous examples of pattern recognition in machine learning, with numerous applications in security, access control, and law enforcement, among many others. Pattern recognition with classical algorithms requires significant computational resources, especially when dealing with high-resolution images in an extensive database. Quantum algorithms have been shown to improve the efficiency and speed of many computational tasks, and as such, they could also potentially improve the complexity of the face recognition process. Here, we propose a quantum machine learning algorithm for pattern recognition based on quantum principal component analysis (QPCA), and quantum independent component analysis (QICA). A novel quantum algorithm for finding dissimilarity in the faces based on the computation of trace and determinant of a matrix (image) is also proposed. The overall complexity of our pattern recognition algorithm is $O(N\,\log N)$ -- $N$ is the image dimension. As an input to these pattern recognition algorithms, we consider experimental images obtained from quantum imaging techniques with correlated photons, e.g. ``interaction-free'' imaging or ``ghost'' imaging. Interfacing these imaging techniques with our quantum pattern recognition processor provides input images that possess a better signal-to-noise ratio, lower exposures, and higher resolution, thus speeding up the machine learning process further. Our fully quantum pattern recognition system with quantum algorithm and quantum inputs promises a much-improved image acquisition and identification system with potential applications extending beyond face recognition, e.g., in medical imaging for diagnosing sensitive tissues or biology for protein identification.}
\end{abstract}

\date{\today}
\maketitle

\begin{figure*}
\centering
\includegraphics[width=18cm]{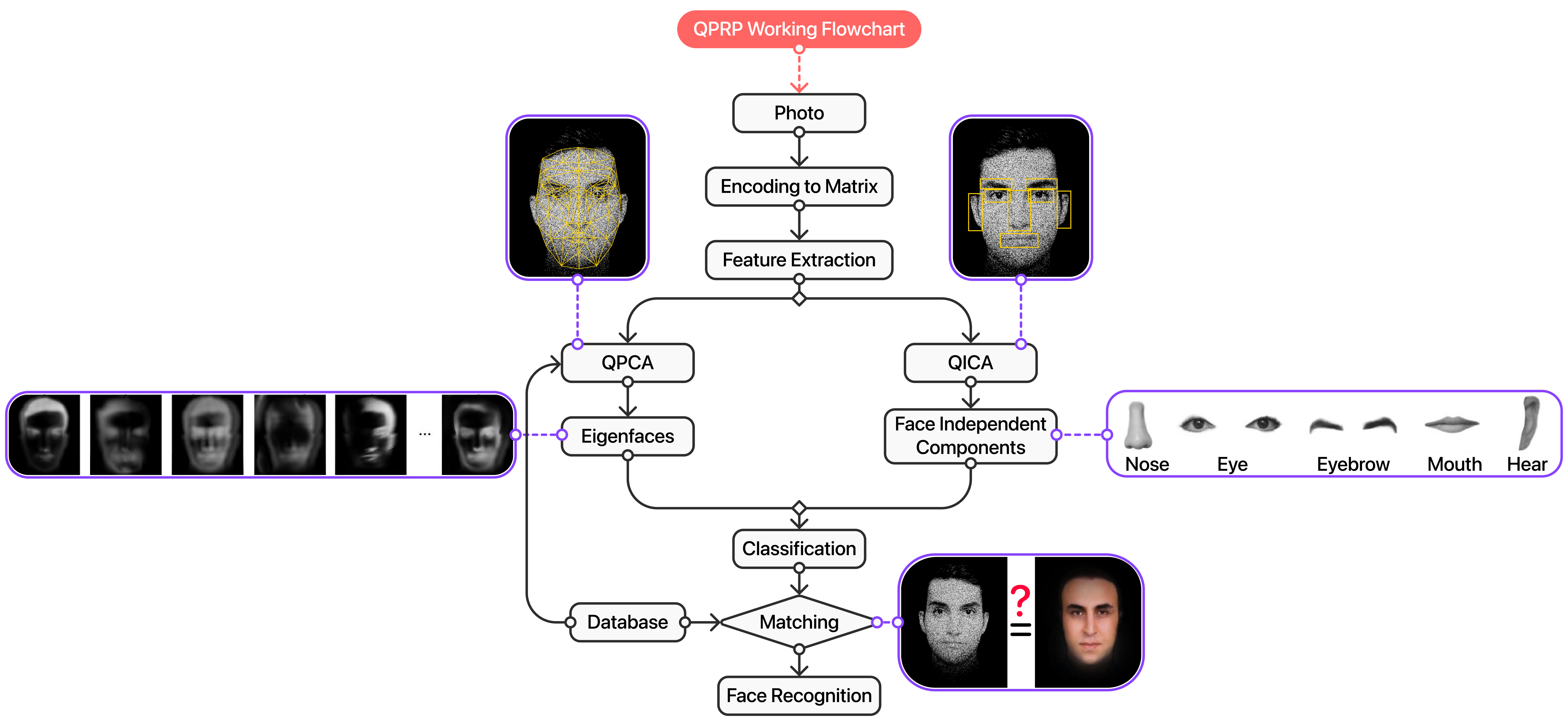}
\caption{\textbf{Flowchart of the quantum algorithm for face recognition.} The quantum algorithm is proposed to be performed in a quantum processor, which we call it quantum pattern recognition processor (QPRP). First the image is converted into matrix form, on which feature extraction algorithms such as quantum principal component analysis (QPCA) or quantum independent component  analysis (QICA) are applied. QPCA extracts the eigenstates (or eigenfaces) of the covariance matrix of the images in the database. The eigenfaces include information like average face, gender (male, female), face direction, brightness, shadows, etc. QICA extracts the independent elements such as eyes, eyebrows, mouth, nose, etc. in a face. The complexity of this stage is O($\log{N})$ -- $N$ is the dimension of th image. Then, the given faces are compared with the faces in the database by using dissimilarity measure based on the log determinant divergence, and the best match among the faces in the database is identified.}\label{Fig:fig1}
\end{figure*}

\section*{Introduction} 
In any intelligent image processing system, there are essentially two main steps: the acquisition of the image and the recognition of the desired patterns. Image acquisition for any pattern recognition method can be performed in multiple ways. For instance, classical sources (incoherent light from thermal radiation or a coherent beam from a laser) or quantum sources (entangled photons obtained from down conversion or squeezed light) can be used to obtain the images. Classical bright field imaging techniques employing the former sources, have the disadvantage of high probe illumination requirement, especially while imaging sensitive samples. Additionally, they are also plagued by the shot noise inherent in the intensities, and the background noise from the environment. Quantum techniques such as quantum illumination, or ghost imaging or even interaction-free imaging, alleviates the problems of background noise, and the probe illumination by utilizing quantum correlations between photon pairs~\cite{shapiro2012physics}. Furthermore, quantum sub-shot noise imaging~\cite{brida2010experimental} and super resolution techniques~\cite{tsang2016quantum} enhance the noise sensitivity and resolution in any images beyond the classical limits. 

As a second important step, pattern recognition in the acquired images is a prominent feature of any intelligent imaging system. Face recognition~\cite{Wright} is one of the branches of pattern recognition, with numerous applications such as face ID verification, passport checks, entrance control, computer access control, criminal investigations, crowd surveillance, and witness face reconstruction~\cite{2D1}, among several others. For face recognition, several classical machine learning algorithms  exist~\cite{hasan2021human}, generally requiring huge computational resources especially when faced with the problem of identification from a large database. Quantum machine learning algorithms employing quantum features such as superposition and entanglement~\cite{lloyd2014quantum,lloyd2014quantum,rebentrost2014quantum,lloyd2016quantum, biamonte2017quantum,Schuld,powell2013quantum,barrios2017role,cardenas2018multiqubit,paneru2020entanglement} promise enhancements in terms of the computing resources and the speed compared to the classical counterparts. Several experimental researches have been done to implement these algorithms~\cite{collins2000nmr,schuch2002implementation,wu2011experimental,takeuchi2000experimental,gulde2003implementation,barz2014two}. In this article, we present a quantum algorithm for face recognition as one of the potential applications of quantum algorithms in machine learning.

The problem of identification of faces from any images generally constitutes different steps (shown in Figure~\ref{Fig:fig1}): creating a database of faces consisting of training and test images, feature extraction using principal component analysis (PCA), linear discriminant analysis (LDA) or independent component analysis (ICA), feature matching using dissimilarity measures, and recognition~\cite{draper2003recognizing}. PCA extracts the eigenstates (or eigenfaces) of the covariance matrix of the images in the database, including information like average face, gender (male or female), face direction, brightness, shadows, etc. ICA, however, extracts the independent elements such as eyes, eyebrows, mouth, nose, etc. in a face. Quantum algorithms which provide speedup for PCA and ICA have already been proposed~\cite{lloyd2014quantum, harrow2009quantum}. Here, we focus on three main steps: (1) Quantum Principle Component Analysis (QPCA) ~\cite{lloyd2014quantum}, (2) Quantum Independent Component Analysis (QICA)~\cite{harrow2009quantum}, and (3) Dissimilarity measures (i.e., face matching), to develop a quantum algorithm for face recognition. In what follows, we present a quantum algorithm for dissimilarity measures for face matching with speedup. This is based on a quantum algorithm to compute the log determinant divergence using both the determinant and the trace of a matrix. Our algorithm combined with the inputs obtained from quantum imaging techniques provides a fully intelligent pattern identification system, with the joint benefit of the low-dose and higher resolution of quantum imaging methods, and the speedup and efficiency of the quantum algorithms.
Figure~\ref{Fig:fig1} shows the flowchart of the quantum algorithm for the pattern identification. 

\begin{figure*}
\centering
\includegraphics[width=18cm]{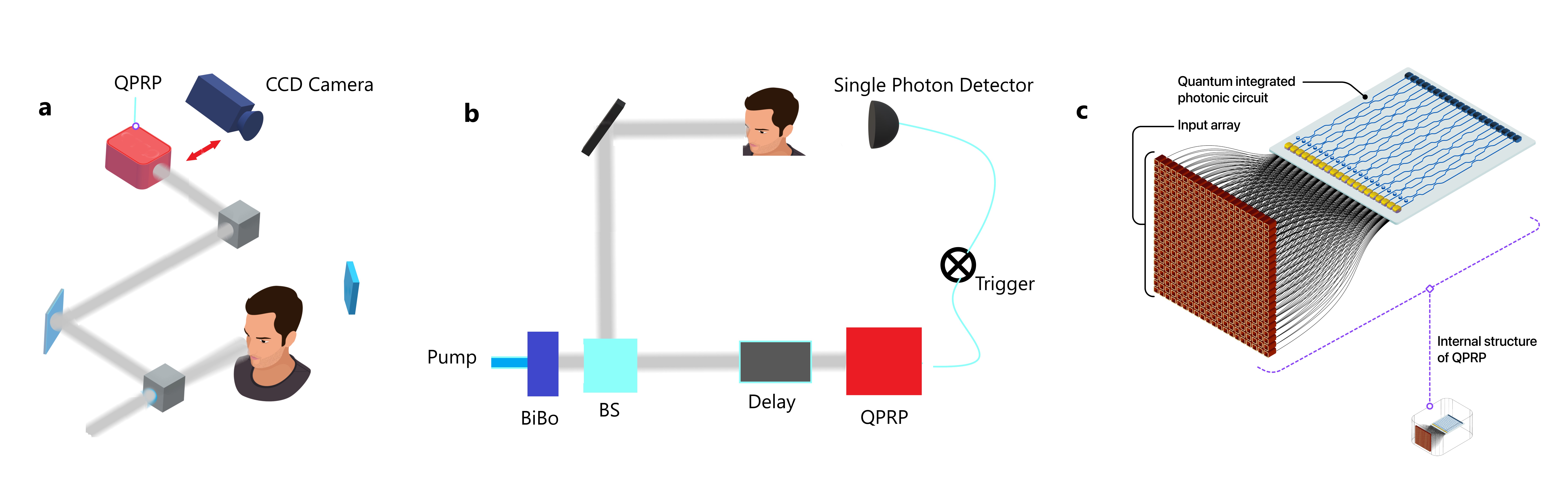}
\caption{\textbf{Intelligent pattern recognition in quantum imaging}. Data from quantum imaging methods such as (\textbf{a}) Interaction Free Imaging and (\textbf{b}) Ghost Imaging act as an input to (\textbf{c}) Quantum Pattern Recognition Processor (QPRP). The latter, i.e., QPRP, applies quantum machine learning to find the patterns in the database.}
\label{Fig:fig2}
\end{figure*}
\section*{Quantum Face Recognition}
Classical algorithms are unable to process quantum data directly. During the conversion of the quantum states (qudits) to classical data (bits), most of the information is lost in the measurement process, due to the ``collapse'' of the wavefunction. Although techniques such as quantum state tomography implemented on unlimited ensemble of the states can be used to fully reconstruct the quantum states from classical projections, these processes are generally complex and expensive. Therefore, the optimal input to our quantum algorithms, would be the quantum states directly obtained from quantum processes, for example, quantum imaging methods, or from a quantum memory, without performing a strong measurement on the wavefunction. 

Photonic quantum memories~\cite{lvovsky2009optical}, allowing storage and on-demand retrieval of quantum states of light, is one of the key components for the realization of quantum optical pattern-recognition technology. Quantum memories essentially form a quantum database for the matching stage in the recognition process. With the state-of-art quantum memories, the possibility of storing hundreds of spatial modes has already been shown in experimental studies using atomic-cold gases~\cite{parniak2017wavevector,pu2017experimental}. Furthermore, using solid-state atomic memories, it is possible to simultaneously store hundreds of photonic quantum states in distinct temporal modes, thus allowing us to store patterns scanned at separate times~\cite{bonarota2011highly,tang2015storage}. In addition, optically accessible spin-states of certain atomic systems can reach several hours of coherence time~\cite{zhong2015optically}. A very recent experimental demonstration reports one-hour memory lifetime for light storage, showing the feasibility of long-lived photonic quantum memory devices~\cite{ma2021one}.
Atomic memory approaches have also been shown to reach high retrieval efficiencies up to $92\%$~\cite{hsiao2018highly} and high fidelities above $99\%$~\cite{liu2020reliable}. 
However, an implementation with all of the aforementioned properties still  remains as a challenge in developing a practical quantum database memory. 

\begin{figure*}
\centering
\includegraphics[width=18cm]{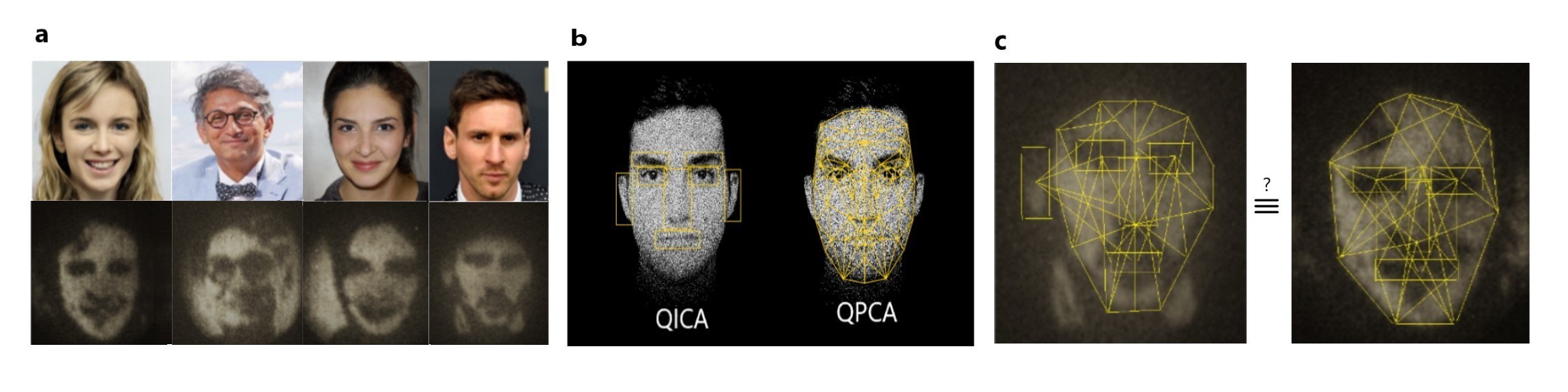}
\caption{\textbf{Face recognition in ghost images.} (\textbf{a}) Images of the original human faces (top) and the corresponding experimental ghost images (bottom) obtained in a ghost imaging setup. A femtosecond laser is used to generate spatially entangled photon pairs. One of the photons illuminates a spatial light modulator, which imprints different images onto the photon, and can act as a trigger for the other photon that was detected by an intensified CCD camera. (\textbf{b}) Quantum Independent Component Analysis (QICA), and Quantum Principal Component Analysis (QPCA), of the faces to detect the independent components, and principal features in the faces. (\textbf{c}) Dissimilarity measure between the ghost images with the images in the database for their identification.}
\label{Fig:fig3}
\end{figure*}

Quantum techniques such as quantum ghost imaging~\cite{morris2015imaging}, quantum lithography~\cite{boto2000quantum}, or quantum sensing~\cite{israel2014supersensitive}, when appropriately interfaced with photonic quantum processors, for example an array of optical fibers connected to an integrated quantum photonic circuit, can also act as inputs to our algorithms (see Figure ~\ref{Fig:fig2}). Here for the case of our face recognition algorithm, we assume that the input images are acquired by quantum ghost imaging ~\cite{morris2015imaging}. Ghost imaging exploits the spatial correlations between photon pairs generated through a nonlinear process called spontaneous parametric down-conversion (SPDC). Since the images are obtained by triggering the shutter in order to capture only the ``coincident'' photon pairs, the level of background noise is significantly reduced, along with a reduction in probe illumination. In a variation of this technique using non degenerate photon pairs, the image detection and sample interaction can happen at different wavelengths, which can be useful when imaging sensitive tissues when limited in detection technologies~\cite{chan2009two}. Combining quantum detection techniques such as interaction-free measurement with ghost imaging, the illumination level required for the same levels of Signal to Noise ratio (SNR) in images~\cite{zhang2019interaction} is further reduced significantly. Figure~\ref{Fig:fig3} shows some of the images of human faces obtained in a quantum ghost imaging setup, where spatially correlated photon pairs (namely signal and idler), are generated by pumping a BiBO crystal with pump photons. Phase holograms placed in a Spatial Light Modulator, a liquid crystal device, created by superimposing the human faces with a diffraction grating acts as an object for the signal photon, while the idler photon passes to the Intensified Charged Coupled Devices (ICCD) camera via a delay line. The images are obtained by triggering the ICCD shutter with the signal photons detected through a Single Photon Avalanche Diode (SPAD) detector -- see Supplementary Information (SI) for the detail of the experimental setup.

\subsection{Quantum principal component analysis (QPCA)} We have now the input images either retrieved from a quantum memory or directly as outputs from a quantum imaging setup. The pattern recognition processor applies Quantum Principal Component Analysis (QPCA)~\cite{lloyd2014quantum,kopczyk2018quantum} to extract the principal eigenvectors of the covariance matrix $C_X$, formed by the set of the training images.

Let us consider a set of  $N$-dimensional training images (or faces), $\{\vert {x}^{(1)} \rangle,\ldots, \vert {x}^{(M)} \rangle\}$. Here, $\vert {x}^{(i)} \rangle$ is the i-th training image, which is given by,
\begin{equation}\label{eq:imagevectors}
    \vert {x}^{(i)} \rangle =\sum_{q =1}^{N} x_q^{(i)} \vert {\psi}^{(i)}_q \rangle,
\end{equation}
where $x_q^{(i)}$ are the components, and $\vert {\psi}^{(i)}_q \rangle$ are the basis kets. The covariance matrix ${C}_X$ can be formed as a sum over $M$ training faces~\cite{kopczyk2018quantum},
\begin{equation}\label{eq:covariancematrix}
    {C}_X = \frac{1}{M} \sum_{i=1}^{M} \vert { x}^{(i)} \rangle \langle { x}^{(i)} \vert.
\end{equation}
The next step is to exponentiate the covariance matrix ${C}_X$, so that we can use the Quantum Phase Estimation (QPE) subroutine for finding the eigenvectors and eigenvalues. It has been shown that the exponentiation of the covariance matrix, i.e., $e^{-i{C}_X\,t}$, can be performed in $\mathcal{O}(\log{}N)$ time~\cite{lloyd2014quantum}.\\

In QPCA algorithm, for the phase estimation subroutine, we apply the operator ${U} = e^{-i{C}_X\, t}$ on ${C}_X$ ~\cite{kopczyk2018quantum}. The action of ${ U}$ on one of the states $\vert  x^{(i)}\rangle $ in ${ C}_X$ is:
\begin{align}
    e^{-i{C}_X\,t} \vert x^{(i)} \rangle \rightarrow  \sum_{j=1}^{M} c^{(ij)} \vert {\phi}^j \rangle,
\end{align}
where $\vert {\phi}^{j} \rangle$'s are the eigenvectors of ${C}_X$, and $c^{ij}=e^{-i\tilde\lambda_c^{(j)}t}\langle \phi^j \vert {x}^{(i)} \rangle$ in which $\tilde\lambda_c^{(j)}={\left(2\pi\lambda_c^{(j)}t\right)}/{2^n}$ where $\lambda_c^{(j)}$'s are the corresponding estimated eigenvalues of ${C}_X$ with precision $n$~\cite{lloyd2014quantum, kopczyk2018quantum}.\\

In order to obtain the principal eigenfaces (the eigenvectors of the covariance matrix with larger eigenvalues), we define a score $s^{(ij)}$, which is the projection of an eigenvector $\vert \phi^{(j)} \rangle$ on a training vector $\vert x^{(i)} \rangle$,
\begin{eqnarray}
    s^{(ij)}=\langle x^i \vert \phi^j \rangle=\sum_{q=1}^{N}x^{(i)}_q \phi^{(j)}_q
    ,
\end{eqnarray}
where $\phi^{(j)}_n$ are the components of the eigenvector $\vert \phi^j \rangle$. The eigenvectors corresponding to the $r$ highest scores are the principal components (or eigenfaces). Each face can be expanded in terms of the $r$ eigenfaces (principal components) but with different weights $\omega_j'$s as follows 
\begin{equation}
    \vert \text{Face}^{(i)} \rangle = \sum_{j=1}^{r}  \omega_j \vert {\mathbf \phi}^j \rangle.
\end{equation}
The ``mean image'' is the eigenface corresponding to the largest eigenvalue of $C_X$. The QPCA algorithm is efficient for the case $r\ll N$~\cite{kopczyk2018quantum}.\\

\subsection{Quantum independent component analysis (QICA)}
In classical machine learning, Independent Component Analysis (ICA) is performed to decompose an observed signal into a linear combination of unknown independent signals~\cite{draper2003recognizing}. Similar to the PCA, the ICA finds a new basis to represent the data, however with a different goal. We assume that there is a data set of faces $s \in {R}^d$ that is a collection of $d$ independent elements in the face such as nose, eye, eyebrow, mouth, etc. Each image observed through a camera can be expressed as ${x}={F}\cdot s$, where ${F}$ is a mixing matrix of the independent face elements. Repeated observation gives us a dataset ${x}$ as $\{x^{(i)},\ldots,x^{(M)}\}$, and ICA estimates the independent sources $s^{(i)}$ that had generated the face.
We let ${W}={F}^{-1}$ which is the unmixing matrix and solve the linear systems of equations $s^{(i)}={W}\,x^{(i)}$ for estimating the independent elements of the face. We should note here that $s^{(i)}$ is a d-dimensional vector and $s^{(i)}_j$ is the data of element $j$. Similarly, $x^{(i)}$ is an d-dimensional vector, and $x^{(i)}_j$ is the observed (or recorded) element $j$ by camera. The ICA can be exponentially speedup via a quantum algorithm for sparse matrices, with the Harrow-Hassidim-Lloyd (HHL) algorithm~\cite{harrow2009quantum}, which is used to solve linear systems of equations optimally with $O(\log{}N)$. For comparison, classically it takes a time $O(N^3)$ to be solved via the Gauss elimination, and approximately $O(N\sqrt{\kappa})$ via iterative methods~\cite{harrow2009quantum} for a sparse matrix of size $N\times N$, with $\kappa$ being the ratio between the greatest and the smallest eigenvalue.

\subsection{Pattern matching: Comparing Faces}
As important details of a face are obtained either by using QPCA or QICA, each face is represented in the form of a sparse matrix in which non-important elements are set to zero. The last and important step of the algorithm is comparing the face patterns to recognize the target face.
Pattern matching algorithms investigate exact matches in the input with pre-existing patterns in the database. In fact, the problem here is comparing matrices with each other. The evaluation of matching between matrices (or face patterns) can be done by using ``dissimilarity'' \cite{cichocki2015log} measures that calculate the ``distance'' between the matrices.  The lower the values of the dissimilarity/distance measures, more similar the matrices, with the fully matched matrices having a zero distance. One such distance measure used to compare two matrices $X$ and $Y$ is called the ``Log-determinant divergence''~\cite{cichocki2015log,dhillon2008matrix} defined as,
\begin{equation}
    D(X,Y)=\text{Tr}\left(X\cdot Y^{-1}\right)-\log \det\left(X\cdot Y^{-1}\right)-N,
\end{equation}
where $N$ is the dimension of the matrices. When $D=0$, the matrices $X$ and $Y$ are completely matched, and higher the distance value the more different are the matrices. The least value among the all distance values identifies the best match and consequently recognizes the face. As it is seen in the distance formula, it is a benefit to be able to calculate the trace and the determinants of matrices with speedup to expedite the distance calculation. In the following, we propose quantum algorithms for computation of the determinant and the trace of a sparse matrix.\newline

\noindent\textbf{Quantum Computation of Sparse Matrix Determinants and Trace:} To obtain a measure of dissimilarity between two matrices we need to calculate the determinant and the trace of the sparse matrix $A = X\cdot Y^{-1}$. First we calculate $Y^{-1}$ using the HHL algorithm~\cite{harrow2009quantum} and obtain $A$ by multiplying it with $X$.  
We then apply the Quantum Phase Estimation (QPE) subroutine, which consists of a quantum Fourier transform (QFT) followed by a controlled Unitary ($\text{CU}$) operation, with $U = e^{-iA\,t}$, and a inverse quantum Fourier transform. We then apply a controlled Rotation operation followed by the inverse Quantum Phase Estimation (QPE) subroutine. At the end we have a multiplication operator $\Pi$ which finally gives us the product of the eigenvalues -- the algorithm steps are explained in more detail in the Supplementary Information. The running time of the algorithm up to the third step, i.e. applying the controlled-U operator, is $O(\log{}N(s^2 \kappa^2 /\epsilon))$~\cite{harrow2009quantum}, where
$s$ is the sparsity, $\kappa$ is the ratio of largest eigenvalue to the smallest eigenvalue of $A$, and $\epsilon$ is the acceptable error. Additionally, the multiplication operation in the last step can be performed in time $O(\log{} N)$ and the algorithm should run $N$ times. Therefore, the overall complexity of the algorithm is $O\left(N\,{\log{}}N(1+ s^2 \kappa^2 / \epsilon)\right) $, which is much faster than the classical ones (see Table~\ref{Table1}).
\begin{table}[h]
\caption{A Comparison of complexities between the classical approaches and our quantum approach, current work (CW), for the computation of determinant.}
\centering
\begin{tabular}{c c c c}
    \bf{Approach} & \bf{Method} & \bf{Complexity} & \bf{Ref.}\\
    \hline\hline
    Classic & Laplace & $N^3$ & \cite{bunch1974triangular}\\
    Classic & Gaussian & $N^3$ & \cite{bunch1974triangular}\\
    Classic & Coppersmith-Winogard & $N^{2.373}$ & \cite{williams2014multiplying} \\
    Classic & Wiedemann & $N^{2} \log N$ & \cite{wiedemann1986solving} \\
    Quantum & Our method & $N \log N$ & CW \\
\end{tabular}
\label{Table1}
\end{table}
\newline

In order to compute the trace of the  matrix $A$, an adder quantum algorithm~\cite{ruiz2017quantum} can speedup the computation. The adder operation between two diagonal elements is mainly based on the quantum Fourier transform (QFT), i.e. $\vert \Phi (a) \rangle :=\text{QFT}\,\vert a \rangle= \frac{1}{\sqrt N}\sum_{k=0}^{N-1}e^{i\frac{2\pi a k}{N}}\vert k \rangle$ and the inverse QFT, i.e., $\text{QFT}^{-1}\vert \Phi (a) \rangle=\vert a \rangle$. By continuation of this method sequentially for the all diagonal elements, one can obtain the trace of the matrix. The detail of the adder algorithm and the quantum circuit for the computation of trace is discussed in the Supplementary Information (Figure~\ref{fig:s2} shows the corresponding quantum circuit). The whole process which is based on QFT and QFT$^{-1}$ has a complexity of $O(\log{} N)$.

\begin{table}[h]
\caption {Summary of estimated complexities in quantum face recognition algorithm.}
\centering
\begin{tabular}{c c c c}
    \bf{Method} & \bf{Output} & \bf{Complexity} & \bf{Ref.}\\
    \hline
    \hline
    QPCA & Eigenfaces & $\log N$ & \cite{lloyd2014quantum}\\
    QICA $\&$ HHL & Face components & $\log N$ & \cite{harrow2009quantum}\\
    HHL & Matrix inversion & $\log N$ & \cite{harrow2009quantum} \\
    Our method & Determinant calculation & $N \log N$ & CW \\
    Our method & Trace calculation & $\log N$ & CW \\
    Log-det divergence & Face matching & $N \log N$ & CW \\
    Our method (General) & Face recognition & $N \log N$ & CW 
\end{tabular}
\label{Table2}
\end{table}

QPCA and QICA both have logarithmic complexities, i.e., $O(\log{}N)$. For the calculation of the log determinant divergence, the computation of trace has a complexity of $O(\log{}N)$, while the determinant has complexity of $O(N\log{}N)$). Hence, the overall complexity of the whole algorithm is $O(N\log{} N)$. Table~\ref{Table2} shows a summary of estimated complexities along with the complexity of the general quantum face recognition algorithm.

\section*{Conclusion}
In summary, we propose a new concept of a quantum protocol for 2D face recognition, combining the benefits of quantum imaging in image acquisition with the speedup from the quantum machine learning algorithms. In this concept, we consider images to be obtained via a ghost imaging protocol either as inputs to the quantum memories or as a hardware encoding of quantum information for the photonic pattern recognition processor. Feeding the ``images'' directly from a quantum protocol also eliminates the need for the conversion of classical data to quantum inputs for the processor saving valuable computational resources. The quantum pattern recognition processor then runs an algorithm composed of three main subroutines: (1) quantum  principal components analysis (QPCA), (2) quantum independent component analysis (QICA), and (3) quantum dissimilarity measures for comparing faces. For the QPCA and QICA, we propose slight modifications in the existing algorithms, whereas for finding the dissimilarity measure, we propose a novel algorithm for obtaining the distance between two matrices based upon a metric called log-determinant divergence. Our algorithm obtains the determinant and the trace of the two matrices in $O(N\log{N})$ time -- N is the dimension of the matrix. Complexity analysis shows that all of the three parts have speedup as compared to their classical counterparts, with the overall complexity given by $O(N\log{N})$. Our conceptual protocol provides a framework for an intelligent and fully quantum image recognition system with quantum inputs and a quantum machine learning processor. The joint benefits of the quantum image acquisition and quantum machine learning promises exciting technological developments in the field of image recognition systems.


\begin{thebibliography}{44}%
\makeatletter
\providecommand \@ifxundefined [1]{%
 \@ifx{#1\undefined}
}%
\providecommand \@ifnum [1]{%
 \ifnum #1\expandafter \@firstoftwo
 \else \expandafter \@secondoftwo
 \fi
}%
\providecommand \@ifx [1]{%
 \ifx #1\expandafter \@firstoftwo
 \else \expandafter \@secondoftwo
 \fi
}%
\providecommand \natexlab [1]{#1}%
\providecommand \enquote  [1]{``#1''}%
\providecommand \bibnamefont  [1]{#1}%
\providecommand \bibfnamefont [1]{#1}%
\providecommand \citenamefont [1]{#1}%
\providecommand \href@noop [0]{\@secondoftwo}%
\providecommand \href [0]{\begingroup \@sanitize@url \@href}%
\providecommand \@href[1]{\@@startlink{#1}\@@href}%
\providecommand \@@href[1]{\endgroup#1\@@endlink}%
\providecommand \@sanitize@url [0]{\catcode `\\12\catcode `\$12\catcode
  `\&12\catcode `\#12\catcode `\^12\catcode `\_12\catcode `\%12\relax}%
\providecommand \@@startlink[1]{}%
\providecommand \@@endlink[0]{}%
\providecommand \url  [0]{\begingroup\@sanitize@url \@url }%
\providecommand \@url [1]{\endgroup\@href {#1}{\urlprefix }}%
\providecommand \urlprefix  [0]{URL }%
\providecommand \Eprint [0]{\href }%
\providecommand \doibase [0]{http://dx.doi.org/}%
\providecommand \selectlanguage [0]{\@gobble}%
\providecommand \bibinfo  [0]{\@secondoftwo}%
\providecommand \bibfield  [0]{\@secondoftwo}%
\providecommand \translation [1]{[#1]}%
\providecommand \BibitemOpen [0]{}%
\providecommand \bibitemStop [0]{}%
\providecommand \bibitemNoStop [0]{.\EOS\space}%
\providecommand \EOS [0]{\spacefactor3000\relax}%
\providecommand \BibitemShut  [1]{\csname bibitem#1\endcsname}%
\let\auto@bib@innerbib\@empty
\bibitem [{\citenamefont {Shapiro}\ and\ \citenamefont
  {Boyd}(2012)}]{shapiro2012physics}%
  \BibitemOpen
  \bibfield  {author} {\bibinfo {author} {\bibfnamefont {Jeffrey~H}\
  \bibnamefont {Shapiro}}\ and\ \bibinfo {author} {\bibfnamefont {Robert~W}\
  \bibnamefont {Boyd}},\ }\bibfield  {title} {\enquote {\bibinfo {title} {The
  physics of ghost imaging},}\ }\href@noop {} {\bibfield  {journal} {\bibinfo
  {journal} {Quantum Information Processing}\ }\textbf {\bibinfo {volume}
  {11}},\ \bibinfo {pages} {949--993} (\bibinfo {year} {2012})}\BibitemShut
  {NoStop}%
\bibitem [{\citenamefont {Brida}\ \emph {et~al.}(2010)\citenamefont {Brida},
  \citenamefont {Genovese},\ and\ \citenamefont
  {Berchera}}]{brida2010experimental}%
  \BibitemOpen
  \bibfield  {author} {\bibinfo {author} {\bibfnamefont {Giorgio}\ \bibnamefont
  {Brida}}, \bibinfo {author} {\bibfnamefont {Marco}\ \bibnamefont {Genovese}},
  \ and\ \bibinfo {author} {\bibfnamefont {I~Ruo}\ \bibnamefont {Berchera}},\
  }\bibfield  {title} {\enquote {\bibinfo {title} {Experimental realization of
  sub-shot-noise quantum imaging},}\ }\href@noop {} {\bibfield  {journal}
  {\bibinfo  {journal} {Nature Photonics}\ }\textbf {\bibinfo {volume} {4}},\
  \bibinfo {pages} {227--230} (\bibinfo {year} {2010})}\BibitemShut {NoStop}%
\bibitem [{\citenamefont {Tsang}\ \emph {et~al.}(2016)\citenamefont {Tsang},
  \citenamefont {Nair},\ and\ \citenamefont {Lu}}]{tsang2016quantum}%
  \BibitemOpen
  \bibfield  {author} {\bibinfo {author} {\bibfnamefont {Mankei}\ \bibnamefont
  {Tsang}}, \bibinfo {author} {\bibfnamefont {Ranjith}\ \bibnamefont {Nair}}, \
  and\ \bibinfo {author} {\bibfnamefont {Xiao-Ming}\ \bibnamefont {Lu}},\
  }\bibfield  {title} {\enquote {\bibinfo {title} {Quantum theory of
  superresolution for two incoherent optical point sources},}\ }\href@noop {}
  {\bibfield  {journal} {\bibinfo  {journal} {Physical Review X}\ }\textbf
  {\bibinfo {volume} {6}},\ \bibinfo {pages} {031033} (\bibinfo {year}
  {2016})}\BibitemShut {NoStop}%
\bibitem [{\citenamefont {J.~Wright}(2009)}]{Wright}%
  \BibitemOpen
  \bibfield  {author} {\bibinfo {author} {\bibfnamefont {et~al.}\ \bibnamefont
  {J.~Wright}},\ }\bibfield  {title} {\enquote {\bibinfo {title} {Robust face
  recognition via sparse representation},}\ }\href@noop {} {\bibfield
  {journal} {\bibinfo  {journal} {IEEE Trans. Patt. Anal. Mach. Intell.}\
  }\textbf {\bibinfo {volume} {31}},\ \bibinfo {pages} {210} (\bibinfo {year}
  {2009})}\BibitemShut {NoStop}%
\bibitem [{\citenamefont {O.~Toygar}(2003)}]{2D1}%
  \BibitemOpen
  \bibfield  {author} {\bibinfo {author} {\bibfnamefont {A.~Acan}\ \bibnamefont
  {O.~Toygar}},\ }\bibfield  {title} {\enquote {\bibinfo {title} {Face
  recognition using pca, lda and ica approaches on colored images},}\
  }\href@noop {} {\bibfield  {journal} {\bibinfo  {journal} {J. Elect. Elect.
  Eng.}\ }\textbf {\bibinfo {volume} {3}},\ \bibinfo {pages} {735} (\bibinfo
  {year} {2003})}\BibitemShut {NoStop}%
\bibitem [{\citenamefont {Hasan}\ \emph {et~al.}(2021)\citenamefont {Hasan},
  \citenamefont {Ahsan}, \citenamefont {Newaz}, \citenamefont {Lee} \emph
  {et~al.}}]{hasan2021human}%
  \BibitemOpen
  \bibfield  {author} {\bibinfo {author} {\bibfnamefont {Md~Khaled}\
  \bibnamefont {Hasan}}, \bibinfo {author} {\bibfnamefont {Md}~\bibnamefont
  {Ahsan}}, \bibinfo {author} {\bibfnamefont {SH}~\bibnamefont {Newaz}},
  \bibinfo {author} {\bibfnamefont {Gyu~Myoung}\ \bibnamefont {Lee}},  \emph
  {et~al.},\ }\bibfield  {title} {\enquote {\bibinfo {title} {Human face
  detection techniques: A comprehensive review and future research
  directions},}\ }\href@noop {} {\bibfield  {journal} {\bibinfo  {journal}
  {Electronics}\ }\textbf {\bibinfo {volume} {10}},\ \bibinfo {pages} {2354}
  (\bibinfo {year} {2021})}\BibitemShut {NoStop}%
\bibitem [{\citenamefont {Lloyd}\ \emph {et~al.}(2014)\citenamefont {Lloyd},
  \citenamefont {Mohseni},\ and\ \citenamefont
  {Rebentrost}}]{lloyd2014quantum}%
  \BibitemOpen
  \bibfield  {author} {\bibinfo {author} {\bibfnamefont {Seth}\ \bibnamefont
  {Lloyd}}, \bibinfo {author} {\bibfnamefont {Masoud}\ \bibnamefont {Mohseni}},
  \ and\ \bibinfo {author} {\bibfnamefont {Patrick}\ \bibnamefont
  {Rebentrost}},\ }\bibfield  {title} {\enquote {\bibinfo {title} {Quantum
  principal component analysis},}\ }\href@noop {} {\bibfield  {journal}
  {\bibinfo  {journal} {Nature Physics}\ }\textbf {\bibinfo {volume} {10}},\
  \bibinfo {pages} {631--633} (\bibinfo {year} {2014})}\BibitemShut {NoStop}%
\bibitem [{\citenamefont {Rebentrost}\ \emph {et~al.}(2014)\citenamefont
  {Rebentrost}, \citenamefont {Mohseni},\ and\ \citenamefont
  {Lloyd}}]{rebentrost2014quantum}%
  \BibitemOpen
  \bibfield  {author} {\bibinfo {author} {\bibfnamefont {Patrick}\ \bibnamefont
  {Rebentrost}}, \bibinfo {author} {\bibfnamefont {Masoud}\ \bibnamefont
  {Mohseni}}, \ and\ \bibinfo {author} {\bibfnamefont {Seth}\ \bibnamefont
  {Lloyd}},\ }\bibfield  {title} {\enquote {\bibinfo {title} {Quantum support
  vector machine for big data classification},}\ }\href@noop {} {\bibfield
  {journal} {\bibinfo  {journal} {Physical review letters}\ }\textbf {\bibinfo
  {volume} {113}},\ \bibinfo {pages} {130503} (\bibinfo {year}
  {2014})}\BibitemShut {NoStop}%
\bibitem [{\citenamefont {Lloyd}\ \emph {et~al.}(2016)\citenamefont {Lloyd},
  \citenamefont {Garnerone},\ and\ \citenamefont {Zanardi}}]{lloyd2016quantum}%
  \BibitemOpen
  \bibfield  {author} {\bibinfo {author} {\bibfnamefont {Seth}\ \bibnamefont
  {Lloyd}}, \bibinfo {author} {\bibfnamefont {Silvano}\ \bibnamefont
  {Garnerone}}, \ and\ \bibinfo {author} {\bibfnamefont {Paolo}\ \bibnamefont
  {Zanardi}},\ }\bibfield  {title} {\enquote {\bibinfo {title} {Quantum
  algorithms for topological and geometric analysis of data},}\ }\href@noop {}
  {\bibfield  {journal} {\bibinfo  {journal} {Nature communications}\ }\textbf
  {\bibinfo {volume} {7}},\ \bibinfo {pages} {1--7} (\bibinfo {year}
  {2016})}\BibitemShut {NoStop}%
\bibitem [{\citenamefont {Biamonte}\ \emph {et~al.}(2017)\citenamefont
  {Biamonte}, \citenamefont {Wittek}, \citenamefont {Pancotti}, \citenamefont
  {Rebentrost}, \citenamefont {Wiebe},\ and\ \citenamefont
  {Lloyd}}]{biamonte2017quantum}%
  \BibitemOpen
  \bibfield  {author} {\bibinfo {author} {\bibfnamefont {Jacob}\ \bibnamefont
  {Biamonte}}, \bibinfo {author} {\bibfnamefont {Peter}\ \bibnamefont
  {Wittek}}, \bibinfo {author} {\bibfnamefont {Nicola}\ \bibnamefont
  {Pancotti}}, \bibinfo {author} {\bibfnamefont {Patrick}\ \bibnamefont
  {Rebentrost}}, \bibinfo {author} {\bibfnamefont {Nathan}\ \bibnamefont
  {Wiebe}}, \ and\ \bibinfo {author} {\bibfnamefont {Seth}\ \bibnamefont
  {Lloyd}},\ }\bibfield  {title} {\enquote {\bibinfo {title} {Quantum machine
  learning},}\ }\href@noop {} {\bibfield  {journal} {\bibinfo  {journal}
  {Nature}\ }\textbf {\bibinfo {volume} {549}},\ \bibinfo {pages} {195--202}
  (\bibinfo {year} {2017})}\BibitemShut {NoStop}%
\bibitem [{\citenamefont {M.~Schuld}(1993)}]{Schuld}%
  \BibitemOpen
  \bibfield  {author} {\bibinfo {author} {\bibfnamefont {F.~Petruccione}\
  \bibnamefont {M.~Schuld}},\ }\href@noop {} {\emph {\bibinfo {title}
  {Supervised Learning with Quantum Computers}}},\ \bibinfo {edition} {3rd}\
  ed.,\ \bibinfo {series} {10}, Vol.~\bibinfo {volume} {4}\ (\bibinfo
  {publisher} {Springer International Publishing},\ \bibinfo {address} {The
  address},\ \bibinfo {year} {1993})\BibitemShut {NoStop}%
\bibitem [{\citenamefont {Powell}(2013)}]{powell2013quantum}%
  \BibitemOpen
  \bibfield  {author} {\bibinfo {author} {\bibfnamefont {Devin}\ \bibnamefont
  {Powell}},\ }\bibfield  {title} {\enquote {\bibinfo {title} {Quantum boost
  for artificial intelligence},}\ }\href@noop {} {\bibfield  {journal}
  {\bibinfo  {journal} {Nature News}\ } (\bibinfo {year} {2013})}\BibitemShut
  {NoStop}%
\bibitem [{\citenamefont {Barrios}\ \emph {et~al.}(2017)\citenamefont
  {Barrios}, \citenamefont {Albarr{\'a}n-Arriagada}, \citenamefont
  {C{\'a}rdenas-L{\'o}pez}, \citenamefont {Romero},\ and\ \citenamefont
  {Retamal}}]{barrios2017role}%
  \BibitemOpen
  \bibfield  {author} {\bibinfo {author} {\bibfnamefont {G~Alvarado}\
  \bibnamefont {Barrios}}, \bibinfo {author} {\bibfnamefont {F}~\bibnamefont
  {Albarr{\'a}n-Arriagada}}, \bibinfo {author} {\bibfnamefont {FA}~\bibnamefont
  {C{\'a}rdenas-L{\'o}pez}}, \bibinfo {author} {\bibfnamefont {G}~\bibnamefont
  {Romero}}, \ and\ \bibinfo {author} {\bibfnamefont {JC}~\bibnamefont
  {Retamal}},\ }\bibfield  {title} {\enquote {\bibinfo {title} {Role of quantum
  correlations in light-matter quantum heat engines},}\ }\href@noop {}
  {\bibfield  {journal} {\bibinfo  {journal} {Physical Review A}\ }\textbf
  {\bibinfo {volume} {96}},\ \bibinfo {pages} {052119} (\bibinfo {year}
  {2017})}\BibitemShut {NoStop}%
\bibitem [{\citenamefont {C{\'a}rdenas-L{\'o}pez}\ \emph
  {et~al.}(2018)\citenamefont {C{\'a}rdenas-L{\'o}pez}, \citenamefont {Lamata},
  \citenamefont {Retamal},\ and\ \citenamefont
  {Solano}}]{cardenas2018multiqubit}%
  \BibitemOpen
  \bibfield  {author} {\bibinfo {author} {\bibfnamefont {Francisco~A}\
  \bibnamefont {C{\'a}rdenas-L{\'o}pez}}, \bibinfo {author} {\bibfnamefont
  {Lucas}\ \bibnamefont {Lamata}}, \bibinfo {author} {\bibfnamefont
  {Juan~Carlos}\ \bibnamefont {Retamal}}, \ and\ \bibinfo {author}
  {\bibfnamefont {Enrique}\ \bibnamefont {Solano}},\ }\bibfield  {title}
  {\enquote {\bibinfo {title} {Multiqubit and multilevel quantum reinforcement
  learning with quantum technologies},}\ }\href@noop {} {\bibfield  {journal}
  {\bibinfo  {journal} {PloS one}\ }\textbf {\bibinfo {volume} {13}},\ \bibinfo
  {pages} {e0200455} (\bibinfo {year} {2018})}\BibitemShut {NoStop}%
\bibitem [{\citenamefont {Paneru}\ \emph {et~al.}(2020)\citenamefont {Paneru},
  \citenamefont {Cohen}, \citenamefont {Fickler}, \citenamefont {Boyd},\ and\
  \citenamefont {Karimi}}]{paneru2020entanglement}%
  \BibitemOpen
  \bibfield  {author} {\bibinfo {author} {\bibfnamefont {Dilip}\ \bibnamefont
  {Paneru}}, \bibinfo {author} {\bibfnamefont {Eliahu}\ \bibnamefont {Cohen}},
  \bibinfo {author} {\bibfnamefont {Robert}\ \bibnamefont {Fickler}}, \bibinfo
  {author} {\bibfnamefont {Robert~W}\ \bibnamefont {Boyd}}, \ and\ \bibinfo
  {author} {\bibfnamefont {Ebrahim}\ \bibnamefont {Karimi}},\ }\bibfield
  {title} {\enquote {\bibinfo {title} {Entanglement: quantum or classical?}}\
  }\href@noop {} {\bibfield  {journal} {\bibinfo  {journal} {Reports on
  Progress in Physics}\ }\textbf {\bibinfo {volume} {83}},\ \bibinfo {pages}
  {064001} (\bibinfo {year} {2020})}\BibitemShut {NoStop}%
\bibitem [{\citenamefont {Collins}\ \emph {et~al.}(2000)\citenamefont
  {Collins}, \citenamefont {Kim}, \citenamefont {Holton}, \citenamefont
  {Sierzputowska-Gracz},\ and\ \citenamefont {Stejskal}}]{collins2000nmr}%
  \BibitemOpen
  \bibfield  {author} {\bibinfo {author} {\bibfnamefont {David}\ \bibnamefont
  {Collins}}, \bibinfo {author} {\bibfnamefont {Ki~Wook}\ \bibnamefont {Kim}},
  \bibinfo {author} {\bibfnamefont {William~C}\ \bibnamefont {Holton}},
  \bibinfo {author} {\bibfnamefont {Hanna}\ \bibnamefont
  {Sierzputowska-Gracz}}, \ and\ \bibinfo {author} {\bibfnamefont
  {EO}~\bibnamefont {Stejskal}},\ }\bibfield  {title} {\enquote {\bibinfo
  {title} {Nmr quantum computation with indirectly coupled gates},}\
  }\href@noop {} {\bibfield  {journal} {\bibinfo  {journal} {Physical Review
  A}\ }\textbf {\bibinfo {volume} {62}},\ \bibinfo {pages} {022304} (\bibinfo
  {year} {2000})}\BibitemShut {NoStop}%
\bibitem [{\citenamefont {Schuch}\ and\ \citenamefont
  {Siewert}(2002)}]{schuch2002implementation}%
  \BibitemOpen
  \bibfield  {author} {\bibinfo {author} {\bibfnamefont {N}~\bibnamefont
  {Schuch}}\ and\ \bibinfo {author} {\bibfnamefont {J}~\bibnamefont
  {Siewert}},\ }\bibfield  {title} {\enquote {\bibinfo {title} {Implementation
  of the four-bit deutsch--jozsa algorithm with josephson charge qubits},}\
  }\href@noop {} {\bibfield  {journal} {\bibinfo  {journal} {physica status
  solidi (b)}\ }\textbf {\bibinfo {volume} {233}},\ \bibinfo {pages} {482--489}
  (\bibinfo {year} {2002})}\BibitemShut {NoStop}%
\bibitem [{\citenamefont {Wu}\ \emph {et~al.}(2011)\citenamefont {Wu},
  \citenamefont {Li}, \citenamefont {Zheng}, \citenamefont {Luo}, \citenamefont
  {Feng},\ and\ \citenamefont {Peng}}]{wu2011experimental}%
  \BibitemOpen
  \bibfield  {author} {\bibinfo {author} {\bibfnamefont {Zhen}\ \bibnamefont
  {Wu}}, \bibinfo {author} {\bibfnamefont {Jun}\ \bibnamefont {Li}}, \bibinfo
  {author} {\bibfnamefont {Wenqiang}\ \bibnamefont {Zheng}}, \bibinfo {author}
  {\bibfnamefont {Jun}\ \bibnamefont {Luo}}, \bibinfo {author} {\bibfnamefont
  {Mang}\ \bibnamefont {Feng}}, \ and\ \bibinfo {author} {\bibfnamefont
  {Xinhua}\ \bibnamefont {Peng}},\ }\bibfield  {title} {\enquote {\bibinfo
  {title} {Experimental demonstration of the deutsch-jozsa algorithm in
  homonuclear multispin systems},}\ }\href@noop {} {\bibfield  {journal}
  {\bibinfo  {journal} {Physical Review A}\ }\textbf {\bibinfo {volume} {84}},\
  \bibinfo {pages} {042312} (\bibinfo {year} {2011})}\BibitemShut {NoStop}%
\bibitem [{\citenamefont {Takeuchi}(2000)}]{takeuchi2000experimental}%
  \BibitemOpen
  \bibfield  {author} {\bibinfo {author} {\bibfnamefont {Shigeki}\ \bibnamefont
  {Takeuchi}},\ }\bibfield  {title} {\enquote {\bibinfo {title} {Experimental
  demonstration of a three-qubit quantum computation algorithm using a single
  photon and linear optics},}\ }\href@noop {} {\bibfield  {journal} {\bibinfo
  {journal} {Physical review A}\ }\textbf {\bibinfo {volume} {62}},\ \bibinfo
  {pages} {032301} (\bibinfo {year} {2000})}\BibitemShut {NoStop}%
\bibitem [{\citenamefont {Gulde}\ \emph {et~al.}(2003)\citenamefont {Gulde},
  \citenamefont {Riebe}, \citenamefont {Lancaster}, \citenamefont {Becher},
  \citenamefont {Eschner}, \citenamefont {H{\"a}ffner}, \citenamefont
  {Schmidt-Kaler}, \citenamefont {Chuang},\ and\ \citenamefont
  {Blatt}}]{gulde2003implementation}%
  \BibitemOpen
  \bibfield  {author} {\bibinfo {author} {\bibfnamefont {Stephan}\ \bibnamefont
  {Gulde}}, \bibinfo {author} {\bibfnamefont {Mark}\ \bibnamefont {Riebe}},
  \bibinfo {author} {\bibfnamefont {Gavin~PT}\ \bibnamefont {Lancaster}},
  \bibinfo {author} {\bibfnamefont {Christoph}\ \bibnamefont {Becher}},
  \bibinfo {author} {\bibfnamefont {J{\"u}rgen}\ \bibnamefont {Eschner}},
  \bibinfo {author} {\bibfnamefont {Hartmut}\ \bibnamefont {H{\"a}ffner}},
  \bibinfo {author} {\bibfnamefont {Ferdinand}\ \bibnamefont {Schmidt-Kaler}},
  \bibinfo {author} {\bibfnamefont {Isaac~L}\ \bibnamefont {Chuang}}, \ and\
  \bibinfo {author} {\bibfnamefont {Rainer}\ \bibnamefont {Blatt}},\ }\bibfield
   {title} {\enquote {\bibinfo {title} {Implementation of the deutsch--jozsa
  algorithm on an ion-trap quantum computer},}\ }\href@noop {} {\bibfield
  {journal} {\bibinfo  {journal} {Nature}\ }\textbf {\bibinfo {volume} {421}},\
  \bibinfo {pages} {48--50} (\bibinfo {year} {2003})}\BibitemShut {NoStop}%
\bibitem [{\citenamefont {Barz}\ \emph {et~al.}(2014)\citenamefont {Barz},
  \citenamefont {Kassal}, \citenamefont {Ringbauer}, \citenamefont {Lipp},
  \citenamefont {Daki{\'c}}, \citenamefont {Aspuru-Guzik},\ and\ \citenamefont
  {Walther}}]{barz2014two}%
  \BibitemOpen
  \bibfield  {author} {\bibinfo {author} {\bibfnamefont {Stefanie}\
  \bibnamefont {Barz}}, \bibinfo {author} {\bibfnamefont {Ivan}\ \bibnamefont
  {Kassal}}, \bibinfo {author} {\bibfnamefont {Martin}\ \bibnamefont
  {Ringbauer}}, \bibinfo {author} {\bibfnamefont {Yannick~Ole}\ \bibnamefont
  {Lipp}}, \bibinfo {author} {\bibfnamefont {Borivoje}\ \bibnamefont
  {Daki{\'c}}}, \bibinfo {author} {\bibfnamefont {Al{\'a}n}\ \bibnamefont
  {Aspuru-Guzik}}, \ and\ \bibinfo {author} {\bibfnamefont {Philip}\
  \bibnamefont {Walther}},\ }\bibfield  {title} {\enquote {\bibinfo {title} {A
  two-qubit photonic quantum processor and its application to solving systems
  of linear equations},}\ }\href@noop {} {\bibfield  {journal} {\bibinfo
  {journal} {Scientific reports}\ }\textbf {\bibinfo {volume} {4}},\ \bibinfo
  {pages} {1--5} (\bibinfo {year} {2014})}\BibitemShut {NoStop}%
\bibitem [{\citenamefont {Draper}\ \emph {et~al.}(2003)\citenamefont {Draper},
  \citenamefont {Baek}, \citenamefont {Bartlett},\ and\ \citenamefont
  {Beveridge}}]{draper2003recognizing}%
  \BibitemOpen
  \bibfield  {author} {\bibinfo {author} {\bibfnamefont {Bruce~A}\ \bibnamefont
  {Draper}}, \bibinfo {author} {\bibfnamefont {Kyungim}\ \bibnamefont {Baek}},
  \bibinfo {author} {\bibfnamefont {Marian~Stewart}\ \bibnamefont {Bartlett}},
  \ and\ \bibinfo {author} {\bibfnamefont {J~Ross}\ \bibnamefont {Beveridge}},\
  }\bibfield  {title} {\enquote {\bibinfo {title} {Recognizing faces with pca
  and ica},}\ }\href@noop {} {\bibfield  {journal} {\bibinfo  {journal}
  {Computer vision and image understanding}\ }\textbf {\bibinfo {volume}
  {91}},\ \bibinfo {pages} {115--137} (\bibinfo {year} {2003})}\BibitemShut
  {NoStop}%
\bibitem [{\citenamefont {Harrow}\ \emph {et~al.}(2009)\citenamefont {Harrow},
  \citenamefont {Hassidim},\ and\ \citenamefont {Lloyd}}]{harrow2009quantum}%
  \BibitemOpen
  \bibfield  {author} {\bibinfo {author} {\bibfnamefont {Aram~W}\ \bibnamefont
  {Harrow}}, \bibinfo {author} {\bibfnamefont {Avinatan}\ \bibnamefont
  {Hassidim}}, \ and\ \bibinfo {author} {\bibfnamefont {Seth}\ \bibnamefont
  {Lloyd}},\ }\bibfield  {title} {\enquote {\bibinfo {title} {Quantum algorithm
  for linear systems of equations},}\ }\href@noop {} {\bibfield  {journal}
  {\bibinfo  {journal} {Physical review letters}\ }\textbf {\bibinfo {volume}
  {103}},\ \bibinfo {pages} {150502} (\bibinfo {year} {2009})}\BibitemShut
  {NoStop}%
\bibitem [{\citenamefont {Lvovsky}\ \emph {et~al.}(2009)\citenamefont
  {Lvovsky}, \citenamefont {Sanders},\ and\ \citenamefont
  {Tittel}}]{lvovsky2009optical}%
  \BibitemOpen
  \bibfield  {author} {\bibinfo {author} {\bibfnamefont {Alexander~I}\
  \bibnamefont {Lvovsky}}, \bibinfo {author} {\bibfnamefont {Barry~C}\
  \bibnamefont {Sanders}}, \ and\ \bibinfo {author} {\bibfnamefont {Wolfgang}\
  \bibnamefont {Tittel}},\ }\bibfield  {title} {\enquote {\bibinfo {title}
  {Optical quantum memory},}\ }\href@noop {} {\bibfield  {journal} {\bibinfo
  {journal} {Nature photonics}\ }\textbf {\bibinfo {volume} {3}},\ \bibinfo
  {pages} {706--714} (\bibinfo {year} {2009})}\BibitemShut {NoStop}%
\bibitem [{\citenamefont {Parniak}\ \emph {et~al.}(2017)\citenamefont
  {Parniak}, \citenamefont {Dkabrowski}, \citenamefont {Mazelanik},
  \citenamefont {Leszczy{\'n}ski}, \citenamefont {Lipka},\ and\ \citenamefont
  {Wasilewski}}]{parniak2017wavevector}%
  \BibitemOpen
  \bibfield  {author} {\bibinfo {author} {\bibfnamefont {Micha{\l}}\
  \bibnamefont {Parniak}}, \bibinfo {author} {\bibfnamefont {Micha{\l}}\
  \bibnamefont {Dkabrowski}}, \bibinfo {author} {\bibfnamefont {Mateusz}\
  \bibnamefont {Mazelanik}}, \bibinfo {author} {\bibfnamefont {Adam}\
  \bibnamefont {Leszczy{\'n}ski}}, \bibinfo {author} {\bibfnamefont
  {Micha{\l}}\ \bibnamefont {Lipka}}, \ and\ \bibinfo {author} {\bibfnamefont
  {Wojciech}\ \bibnamefont {Wasilewski}},\ }\bibfield  {title} {\enquote
  {\bibinfo {title} {Wavevector multiplexed atomic quantum memory via
  spatially-resolved single-photon detection},}\ }\href@noop {} {\bibfield
  {journal} {\bibinfo  {journal} {Nature communications}\ }\textbf {\bibinfo
  {volume} {8}},\ \bibinfo {pages} {1--9} (\bibinfo {year} {2017})}\BibitemShut
  {NoStop}%
\bibitem [{\citenamefont {Pu}\ \emph {et~al.}(2017)\citenamefont {Pu},
  \citenamefont {Jiang}, \citenamefont {Chang}, \citenamefont {Yang},
  \citenamefont {Li},\ and\ \citenamefont {Duan}}]{pu2017experimental}%
  \BibitemOpen
  \bibfield  {author} {\bibinfo {author} {\bibfnamefont {YF}~\bibnamefont
  {Pu}}, \bibinfo {author} {\bibfnamefont {N}~\bibnamefont {Jiang}}, \bibinfo
  {author} {\bibfnamefont {W}~\bibnamefont {Chang}}, \bibinfo {author}
  {\bibfnamefont {HX}~\bibnamefont {Yang}}, \bibinfo {author} {\bibfnamefont
  {C}~\bibnamefont {Li}}, \ and\ \bibinfo {author} {\bibfnamefont
  {LM}~\bibnamefont {Duan}},\ }\bibfield  {title} {\enquote {\bibinfo {title}
  {Experimental realization of a multiplexed quantum memory with 225
  individually accessible memory cells},}\ }\href@noop {} {\bibfield  {journal}
  {\bibinfo  {journal} {Nature communications}\ }\textbf {\bibinfo {volume}
  {8}},\ \bibinfo {pages} {1--6} (\bibinfo {year} {2017})}\BibitemShut
  {NoStop}%
\bibitem [{\citenamefont {Bonarota}\ \emph {et~al.}(2011)\citenamefont
  {Bonarota}, \citenamefont {Le~Gou{\"e}t},\ and\ \citenamefont
  {Chaneliere}}]{bonarota2011highly}%
  \BibitemOpen
  \bibfield  {author} {\bibinfo {author} {\bibfnamefont {M}~\bibnamefont
  {Bonarota}}, \bibinfo {author} {\bibfnamefont {JL}~\bibnamefont
  {Le~Gou{\"e}t}}, \ and\ \bibinfo {author} {\bibfnamefont {T}~\bibnamefont
  {Chaneliere}},\ }\bibfield  {title} {\enquote {\bibinfo {title} {Highly
  multimode storage in a crystal},}\ }\href@noop {} {\bibfield  {journal}
  {\bibinfo  {journal} {New Journal of Physics}\ }\textbf {\bibinfo {volume}
  {13}},\ \bibinfo {pages} {013013} (\bibinfo {year} {2011})}\BibitemShut
  {NoStop}%
\bibitem [{\citenamefont {Tang}\ \emph {et~al.}(2015)\citenamefont {Tang},
  \citenamefont {Zhou}, \citenamefont {Wang}, \citenamefont {Li}, \citenamefont
  {Liu}, \citenamefont {Hua}, \citenamefont {Zou}, \citenamefont {Wang},
  \citenamefont {He}, \citenamefont {Chen} \emph {et~al.}}]{tang2015storage}%
  \BibitemOpen
  \bibfield  {author} {\bibinfo {author} {\bibfnamefont {Jian-Shun}\
  \bibnamefont {Tang}}, \bibinfo {author} {\bibfnamefont {Zong-Quan}\
  \bibnamefont {Zhou}}, \bibinfo {author} {\bibfnamefont {Yi-Tao}\ \bibnamefont
  {Wang}}, \bibinfo {author} {\bibfnamefont {Yu-Long}\ \bibnamefont {Li}},
  \bibinfo {author} {\bibfnamefont {Xiao}\ \bibnamefont {Liu}}, \bibinfo
  {author} {\bibfnamefont {Yi-Lin}\ \bibnamefont {Hua}}, \bibinfo {author}
  {\bibfnamefont {Yang}\ \bibnamefont {Zou}}, \bibinfo {author} {\bibfnamefont
  {Shuang}\ \bibnamefont {Wang}}, \bibinfo {author} {\bibfnamefont {De-Yong}\
  \bibnamefont {He}}, \bibinfo {author} {\bibfnamefont {Geng}\ \bibnamefont
  {Chen}},  \emph {et~al.},\ }\bibfield  {title} {\enquote {\bibinfo {title}
  {Storage of multiple single-photon pulses emitted from a quantum dot in a
  solid-state quantum memory},}\ }\href@noop {} {\bibfield  {journal} {\bibinfo
   {journal} {Nature communications}\ }\textbf {\bibinfo {volume} {6}},\
  \bibinfo {pages} {1--7} (\bibinfo {year} {2015})}\BibitemShut {NoStop}%
\bibitem [{\citenamefont {Zhong}\ \emph {et~al.}(2015)\citenamefont {Zhong},
  \citenamefont {Hedges}, \citenamefont {Ahlefeldt}, \citenamefont
  {Bartholomew}, \citenamefont {Beavan}, \citenamefont {Wittig}, \citenamefont
  {Longdell},\ and\ \citenamefont {Sellars}}]{zhong2015optically}%
  \BibitemOpen
  \bibfield  {author} {\bibinfo {author} {\bibfnamefont {Manjin}\ \bibnamefont
  {Zhong}}, \bibinfo {author} {\bibfnamefont {Morgan~P}\ \bibnamefont
  {Hedges}}, \bibinfo {author} {\bibfnamefont {Rose~L}\ \bibnamefont
  {Ahlefeldt}}, \bibinfo {author} {\bibfnamefont {John~G}\ \bibnamefont
  {Bartholomew}}, \bibinfo {author} {\bibfnamefont {Sarah~E}\ \bibnamefont
  {Beavan}}, \bibinfo {author} {\bibfnamefont {Sven~M}\ \bibnamefont {Wittig}},
  \bibinfo {author} {\bibfnamefont {Jevon~J}\ \bibnamefont {Longdell}}, \ and\
  \bibinfo {author} {\bibfnamefont {Matthew~J}\ \bibnamefont {Sellars}},\
  }\bibfield  {title} {\enquote {\bibinfo {title} {Optically addressable
  nuclear spins in a solid with a six-hour coherence time},}\ }\href@noop {}
  {\bibfield  {journal} {\bibinfo  {journal} {Nature}\ }\textbf {\bibinfo
  {volume} {517}},\ \bibinfo {pages} {177--180} (\bibinfo {year}
  {2015})}\BibitemShut {NoStop}%
\bibitem [{\citenamefont {Ma}\ \emph {et~al.}(2021)\citenamefont {Ma},
  \citenamefont {Ma}, \citenamefont {Zhou}, \citenamefont {Li},\ and\
  \citenamefont {Guo}}]{ma2021one}%
  \BibitemOpen
  \bibfield  {author} {\bibinfo {author} {\bibfnamefont {Yu}~\bibnamefont
  {Ma}}, \bibinfo {author} {\bibfnamefont {You-Zhi}\ \bibnamefont {Ma}},
  \bibinfo {author} {\bibfnamefont {Zong-Quan}\ \bibnamefont {Zhou}}, \bibinfo
  {author} {\bibfnamefont {Chuan-Feng}\ \bibnamefont {Li}}, \ and\ \bibinfo
  {author} {\bibfnamefont {Guang-Can}\ \bibnamefont {Guo}},\ }\bibfield
  {title} {\enquote {\bibinfo {title} {One-hour coherent optical storage in an
  atomic frequency comb memory},}\ }\href@noop {} {\bibfield  {journal}
  {\bibinfo  {journal} {Nature communications}\ }\textbf {\bibinfo {volume}
  {12}},\ \bibinfo {pages} {1--6} (\bibinfo {year} {2021})}\BibitemShut
  {NoStop}%
\bibitem [{\citenamefont {Hsiao}\ \emph {et~al.}(2018)\citenamefont {Hsiao},
  \citenamefont {Tsai}, \citenamefont {Chen}, \citenamefont {Lin},
  \citenamefont {Hung}, \citenamefont {Lee}, \citenamefont {Chen},
  \citenamefont {Chen}, \citenamefont {Ite},\ and\ \citenamefont
  {Chen}}]{hsiao2018highly}%
  \BibitemOpen
  \bibfield  {author} {\bibinfo {author} {\bibfnamefont {Ya-Fen}\ \bibnamefont
  {Hsiao}}, \bibinfo {author} {\bibfnamefont {Pin-Ju}\ \bibnamefont {Tsai}},
  \bibinfo {author} {\bibfnamefont {Hung-Shiue}\ \bibnamefont {Chen}}, \bibinfo
  {author} {\bibfnamefont {Sheng-Xiang}\ \bibnamefont {Lin}}, \bibinfo {author}
  {\bibfnamefont {Chih-Chiao}\ \bibnamefont {Hung}}, \bibinfo {author}
  {\bibfnamefont {Chih-Hsi}\ \bibnamefont {Lee}}, \bibinfo {author}
  {\bibfnamefont {Yi-Hsin}\ \bibnamefont {Chen}}, \bibinfo {author}
  {\bibfnamefont {Yong-Fan}\ \bibnamefont {Chen}}, \bibinfo {author}
  {\bibfnamefont {A~Yu}\ \bibnamefont {Ite}}, \ and\ \bibinfo {author}
  {\bibfnamefont {Ying-Cheng}\ \bibnamefont {Chen}},\ }\bibfield  {title}
  {\enquote {\bibinfo {title} {Highly efficient coherent optical memory based
  on electromagnetically induced transparency},}\ }\href@noop {} {\bibfield
  {journal} {\bibinfo  {journal} {Physical review letters}\ }\textbf {\bibinfo
  {volume} {120}},\ \bibinfo {pages} {183602} (\bibinfo {year}
  {2018})}\BibitemShut {NoStop}%
\bibitem [{\citenamefont {Liu}\ \emph {et~al.}(2020)\citenamefont {Liu},
  \citenamefont {Zhou}, \citenamefont {Zhu}, \citenamefont {Zheng},
  \citenamefont {Jin}, \citenamefont {Liu}, \citenamefont {Li}, \citenamefont
  {Huang}, \citenamefont {Ma}, \citenamefont {Tu} \emph
  {et~al.}}]{liu2020reliable}%
  \BibitemOpen
  \bibfield  {author} {\bibinfo {author} {\bibfnamefont {Chao}\ \bibnamefont
  {Liu}}, \bibinfo {author} {\bibfnamefont {Zong-Quan}\ \bibnamefont {Zhou}},
  \bibinfo {author} {\bibfnamefont {Tian-Xiang}\ \bibnamefont {Zhu}}, \bibinfo
  {author} {\bibfnamefont {Liang}\ \bibnamefont {Zheng}}, \bibinfo {author}
  {\bibfnamefont {Ming}\ \bibnamefont {Jin}}, \bibinfo {author} {\bibfnamefont
  {Xiao}\ \bibnamefont {Liu}}, \bibinfo {author} {\bibfnamefont {Pei-Yun}\
  \bibnamefont {Li}}, \bibinfo {author} {\bibfnamefont {Jian-Yin}\ \bibnamefont
  {Huang}}, \bibinfo {author} {\bibfnamefont {Yu}~\bibnamefont {Ma}}, \bibinfo
  {author} {\bibfnamefont {Tao}\ \bibnamefont {Tu}},  \emph {et~al.},\
  }\bibfield  {title} {\enquote {\bibinfo {title} {Reliable coherent optical
  memory based on a laser-written waveguide},}\ }\href@noop {} {\bibfield
  {journal} {\bibinfo  {journal} {Optica}\ }\textbf {\bibinfo {volume} {7}},\
  \bibinfo {pages} {192--197} (\bibinfo {year} {2020})}\BibitemShut {NoStop}%
\bibitem [{\citenamefont {Morris}\ \emph {et~al.}(2015)\citenamefont {Morris},
  \citenamefont {Aspden}, \citenamefont {Bell}, \citenamefont {Boyd},\ and\
  \citenamefont {Padgett}}]{morris2015imaging}%
  \BibitemOpen
  \bibfield  {author} {\bibinfo {author} {\bibfnamefont {Peter~A}\ \bibnamefont
  {Morris}}, \bibinfo {author} {\bibfnamefont {Reuben~S}\ \bibnamefont
  {Aspden}}, \bibinfo {author} {\bibfnamefont {Jessica~EC}\ \bibnamefont
  {Bell}}, \bibinfo {author} {\bibfnamefont {Robert~W}\ \bibnamefont {Boyd}}, \
  and\ \bibinfo {author} {\bibfnamefont {Miles~J}\ \bibnamefont {Padgett}},\
  }\bibfield  {title} {\enquote {\bibinfo {title} {Imaging with a small number
  of photons},}\ }\href@noop {} {\bibfield  {journal} {\bibinfo  {journal}
  {Nature communications}\ }\textbf {\bibinfo {volume} {6}},\ \bibinfo {pages}
  {1--6} (\bibinfo {year} {2015})}\BibitemShut {NoStop}%
\bibitem [{\citenamefont {Boto}\ \emph {et~al.}(2000)\citenamefont {Boto},
  \citenamefont {Kok}, \citenamefont {Abrams}, \citenamefont {Braunstein},
  \citenamefont {Williams},\ and\ \citenamefont {Dowling}}]{boto2000quantum}%
  \BibitemOpen
  \bibfield  {author} {\bibinfo {author} {\bibfnamefont {Agedi~N}\ \bibnamefont
  {Boto}}, \bibinfo {author} {\bibfnamefont {Pieter}\ \bibnamefont {Kok}},
  \bibinfo {author} {\bibfnamefont {Daniel~S}\ \bibnamefont {Abrams}}, \bibinfo
  {author} {\bibfnamefont {Samuel~L}\ \bibnamefont {Braunstein}}, \bibinfo
  {author} {\bibfnamefont {Colin~P}\ \bibnamefont {Williams}}, \ and\ \bibinfo
  {author} {\bibfnamefont {Jonathan~P}\ \bibnamefont {Dowling}},\ }\bibfield
  {title} {\enquote {\bibinfo {title} {Quantum interferometric optical
  lithography: exploiting entanglement to beat the diffraction limit},}\
  }\href@noop {} {\bibfield  {journal} {\bibinfo  {journal} {Physical Review
  Letters}\ }\textbf {\bibinfo {volume} {85}},\ \bibinfo {pages} {2733}
  (\bibinfo {year} {2000})}\BibitemShut {NoStop}%
\bibitem [{\citenamefont {Israel}\ \emph {et~al.}(2014)\citenamefont {Israel},
  \citenamefont {Rosen},\ and\ \citenamefont
  {Silberberg}}]{israel2014supersensitive}%
  \BibitemOpen
  \bibfield  {author} {\bibinfo {author} {\bibfnamefont {Yonatan}\ \bibnamefont
  {Israel}}, \bibinfo {author} {\bibfnamefont {Shamir}\ \bibnamefont {Rosen}},
  \ and\ \bibinfo {author} {\bibfnamefont {Yaron}\ \bibnamefont {Silberberg}},\
  }\bibfield  {title} {\enquote {\bibinfo {title} {Supersensitive polarization
  microscopy using noon states of light},}\ }\href@noop {} {\bibfield
  {journal} {\bibinfo  {journal} {Physical review letters}\ }\textbf {\bibinfo
  {volume} {112}},\ \bibinfo {pages} {103604} (\bibinfo {year}
  {2014})}\BibitemShut {NoStop}%
\bibitem [{\citenamefont {Chan}\ \emph {et~al.}(2009)\citenamefont {Chan},
  \citenamefont {O'Sullivan},\ and\ \citenamefont {Boyd}}]{chan2009two}%
  \BibitemOpen
  \bibfield  {author} {\bibinfo {author} {\bibfnamefont {Kam Wai~Clifford}\
  \bibnamefont {Chan}}, \bibinfo {author} {\bibfnamefont {Malcolm~N}\
  \bibnamefont {O'Sullivan}}, \ and\ \bibinfo {author} {\bibfnamefont
  {Robert~W}\ \bibnamefont {Boyd}},\ }\bibfield  {title} {\enquote {\bibinfo
  {title} {Two-color ghost imaging},}\ }\href@noop {} {\bibfield  {journal}
  {\bibinfo  {journal} {Physical Review A}\ }\textbf {\bibinfo {volume} {79}},\
  \bibinfo {pages} {033808} (\bibinfo {year} {2009})}\BibitemShut {NoStop}%
\bibitem [{\citenamefont {Zhang}\ \emph {et~al.}(2019)\citenamefont {Zhang},
  \citenamefont {Sit}, \citenamefont {Bouchard}, \citenamefont {Larocque},
  \citenamefont {Grenapin}, \citenamefont {Cohen}, \citenamefont {Elitzur},
  \citenamefont {Harden}, \citenamefont {Boyd},\ and\ \citenamefont
  {Karimi}}]{zhang2019interaction}%
  \BibitemOpen
  \bibfield  {author} {\bibinfo {author} {\bibfnamefont {Yingwen}\ \bibnamefont
  {Zhang}}, \bibinfo {author} {\bibfnamefont {Alicia}\ \bibnamefont {Sit}},
  \bibinfo {author} {\bibfnamefont {Fr{\'e}d{\'e}ric}\ \bibnamefont
  {Bouchard}}, \bibinfo {author} {\bibfnamefont {Hugo}\ \bibnamefont
  {Larocque}}, \bibinfo {author} {\bibfnamefont {Florence}\ \bibnamefont
  {Grenapin}}, \bibinfo {author} {\bibfnamefont {Eliahu}\ \bibnamefont
  {Cohen}}, \bibinfo {author} {\bibfnamefont {Avshalom~C}\ \bibnamefont
  {Elitzur}}, \bibinfo {author} {\bibfnamefont {James~L}\ \bibnamefont
  {Harden}}, \bibinfo {author} {\bibfnamefont {Robert~W}\ \bibnamefont {Boyd}},
  \ and\ \bibinfo {author} {\bibfnamefont {Ebrahim}\ \bibnamefont {Karimi}},\
  }\bibfield  {title} {\enquote {\bibinfo {title} {Interaction-free
  ghost-imaging of structured objects},}\ }\href@noop {} {\bibfield  {journal}
  {\bibinfo  {journal} {Optics express}\ }\textbf {\bibinfo {volume} {27}},\
  \bibinfo {pages} {2212--2224} (\bibinfo {year} {2019})}\BibitemShut {NoStop}%
\bibitem [{\citenamefont {Kopczyk}(2018)}]{kopczyk2018quantum}%
  \BibitemOpen
  \bibfield  {author} {\bibinfo {author} {\bibfnamefont {Dawid}\ \bibnamefont
  {Kopczyk}},\ }\bibfield  {title} {\enquote {\bibinfo {title} {Quantum machine
  learning for data scientists},}\ }\href@noop {} {\bibfield  {journal}
  {\bibinfo  {journal} {arXiv preprint arXiv:1804.10068}\ } (\bibinfo {year}
  {2018})}\BibitemShut {NoStop}%
\bibitem [{\citenamefont {Cichocki}\ \emph {et~al.}(2015)\citenamefont
  {Cichocki}, \citenamefont {Cruces},\ and\ \citenamefont
  {Amari}}]{cichocki2015log}%
  \BibitemOpen
  \bibfield  {author} {\bibinfo {author} {\bibfnamefont {Andrzej}\ \bibnamefont
  {Cichocki}}, \bibinfo {author} {\bibfnamefont {Sergio}\ \bibnamefont
  {Cruces}}, \ and\ \bibinfo {author} {\bibfnamefont {Shun-ichi}\ \bibnamefont
  {Amari}},\ }\bibfield  {title} {\enquote {\bibinfo {title} {Log-determinant
  divergences revisited: Alpha-beta and gamma log-det divergences},}\
  }\href@noop {} {\bibfield  {journal} {\bibinfo  {journal} {Entropy}\ }\textbf
  {\bibinfo {volume} {17}},\ \bibinfo {pages} {2988--3034} (\bibinfo {year}
  {2015})}\BibitemShut {NoStop}%
\bibitem [{\citenamefont {Dhillon}\ and\ \citenamefont
  {Tropp}(2008)}]{dhillon2008matrix}%
  \BibitemOpen
  \bibfield  {author} {\bibinfo {author} {\bibfnamefont {Inderjit~S}\
  \bibnamefont {Dhillon}}\ and\ \bibinfo {author} {\bibfnamefont {Joel~A}\
  \bibnamefont {Tropp}},\ }\bibfield  {title} {\enquote {\bibinfo {title}
  {Matrix nearness problems with bregman divergences},}\ }\href@noop {}
  {\bibfield  {journal} {\bibinfo  {journal} {SIAM Journal on Matrix Analysis
  and Applications}\ }\textbf {\bibinfo {volume} {29}},\ \bibinfo {pages}
  {1120--1146} (\bibinfo {year} {2008})}\BibitemShut {NoStop}%
\bibitem [{\citenamefont {Bunch}\ and\ \citenamefont
  {Hopcroft}(1974)}]{bunch1974triangular}%
  \BibitemOpen
  \bibfield  {author} {\bibinfo {author} {\bibfnamefont {James~R}\ \bibnamefont
  {Bunch}}\ and\ \bibinfo {author} {\bibfnamefont {John~E}\ \bibnamefont
  {Hopcroft}},\ }\bibfield  {title} {\enquote {\bibinfo {title} {Triangular
  factorization and inversion by fast matrix multiplication},}\ }\href@noop {}
  {\bibfield  {journal} {\bibinfo  {journal} {Mathematics of Computation}\
  }\textbf {\bibinfo {volume} {28}},\ \bibinfo {pages} {231--236} (\bibinfo
  {year} {1974})}\BibitemShut {NoStop}%
\bibitem [{\citenamefont {Williams}(2014)}]{williams2014multiplying}%
  \BibitemOpen
  \bibfield  {author} {\bibinfo {author} {\bibfnamefont {Virginia~Vassilevska}\
  \bibnamefont {Williams}},\ }\bibfield  {title} {\enquote {\bibinfo {title}
  {Multiplying matrices in o (n2. 373) time},}\ }\href@noop {} {\bibfield
  {journal} {\bibinfo  {journal} {preprint}\ } (\bibinfo {year}
  {2014})}\BibitemShut {NoStop}%
\bibitem [{\citenamefont {Wiedemann}(1986)}]{wiedemann1986solving}%
  \BibitemOpen
  \bibfield  {author} {\bibinfo {author} {\bibfnamefont {Douglas}\ \bibnamefont
  {Wiedemann}},\ }\bibfield  {title} {\enquote {\bibinfo {title} {Solving
  sparse linear equations over finite fields},}\ }\href@noop {} {\bibfield
  {journal} {\bibinfo  {journal} {IEEE transactions on information theory}\
  }\textbf {\bibinfo {volume} {32}},\ \bibinfo {pages} {54--62} (\bibinfo
  {year} {1986})}\BibitemShut {NoStop}%
\bibitem [{\citenamefont {Ruiz-Perez}\ and\ \citenamefont
  {Garcia-Escartin}(2017)}]{ruiz2017quantum}%
  \BibitemOpen
  \bibfield  {author} {\bibinfo {author} {\bibfnamefont {Lidia}\ \bibnamefont
  {Ruiz-Perez}}\ and\ \bibinfo {author} {\bibfnamefont {Juan~Carlos}\
  \bibnamefont {Garcia-Escartin}},\ }\bibfield  {title} {\enquote {\bibinfo
  {title} {Quantum arithmetic with the quantum fourier transform},}\
  }\href@noop {} {\bibfield  {journal} {\bibinfo  {journal} {Quantum
  Information Processing}\ }\textbf {\bibinfo {volume} {16}},\ \bibinfo {pages}
  {152} (\bibinfo {year} {2017})}\BibitemShut {NoStop}%
\end{thebibliography}

%

\vspace{0.6 EM}


\noindent\textbf{Acknowledgments:}
\noindent V. S. is very thankful for several helpful discussions with Mikel Sanz and Enrique Solano during his research stay in QUTIS center in Bilbao (Spain) and QuArtist center in Shanghai (China), both leading by Enrique Solano who encouraged the idea to be developed. Also, V. S. is very grateful for useful discussions with Seth Lloyd and Nathan Wiebe during the Quantum Machin Learning and Biomimetic Quantum Technologies conference held in Bilbao. \\

\noindent\textbf{Funding:}
\noindent D. P., M. R. and E. K. acknowledge the support of Ontario's Early Researcher Award (ERA), Canada Research Chairs (CRC), and the European Union's Horizon 2020 Research and Innovation Programme (Q-SORT), grant number 766970. V. S. is grateful for the financial support by the Spanish State Research Agency through BCAM Severo Ochoa excellence accreditation SEV-2017-0718 and BERC 2018-2021 program. S.B. acknowledges funding by the Natural Sciences and Engineering Research Council of Canada (NSERC) through its Discovery Grant, funding and advisory support provided by Alberta Innovates through the Accelerating Innovations into CarE (AICE) -- Concepts Program, and support from Alberta Innovates and NSERC through Advance Grant.\\

\noindent\textbf{Author contributions:} V.S. and D.P. contributed equally to this work. V.S. and E.K. developed the idea and consulted it with D.P., S.B., M.R., and E.S. to design the protocol. V.S., M.G., and M.Ab. developed the algorithm, D.P. performed the experiments under the supervision of E.K. All authors contributed to discussions. V.S., D.P., and E.K. wrote the manuscript, and V.S., S.B., and E.K. revised the final version.
\\

\noindent\textbf{Competing interests:} 
The authors declare that they have no competing interests.
\\

\noindent\textbf{Data and materials availability:} 
All data needed to evaluate the conclusions in the paper are present in the paper
and/or the Supplementary Materials. Additional data related to this paper may be requested from the authors.

\clearpage
\renewcommand{\figurename}{\textbf{Supplementary Figure}}
\setcounter{figure}{0} \renewcommand{\thefigure}{\textbf{S{\arabic{figure}}}}
\setcounter{table}{0} \renewcommand{\thetable}{S\arabic{table}}
\setcounter{section}{0} \renewcommand{\thesection}{S\arabic{section}}
\setcounter{equation}{0} \renewcommand{\theequation}{S\arabic{equation}}
\onecolumngrid

\begin{center}
{\Large Supplementary Information for: \\ Quantum Face Recognition Protocol with Ghost Imaging}
\end{center}

\section{Quantum Imaging}\label{Section:S1}
\begin{figure*}[b]
    \centering
    \includegraphics[scale=0.25]{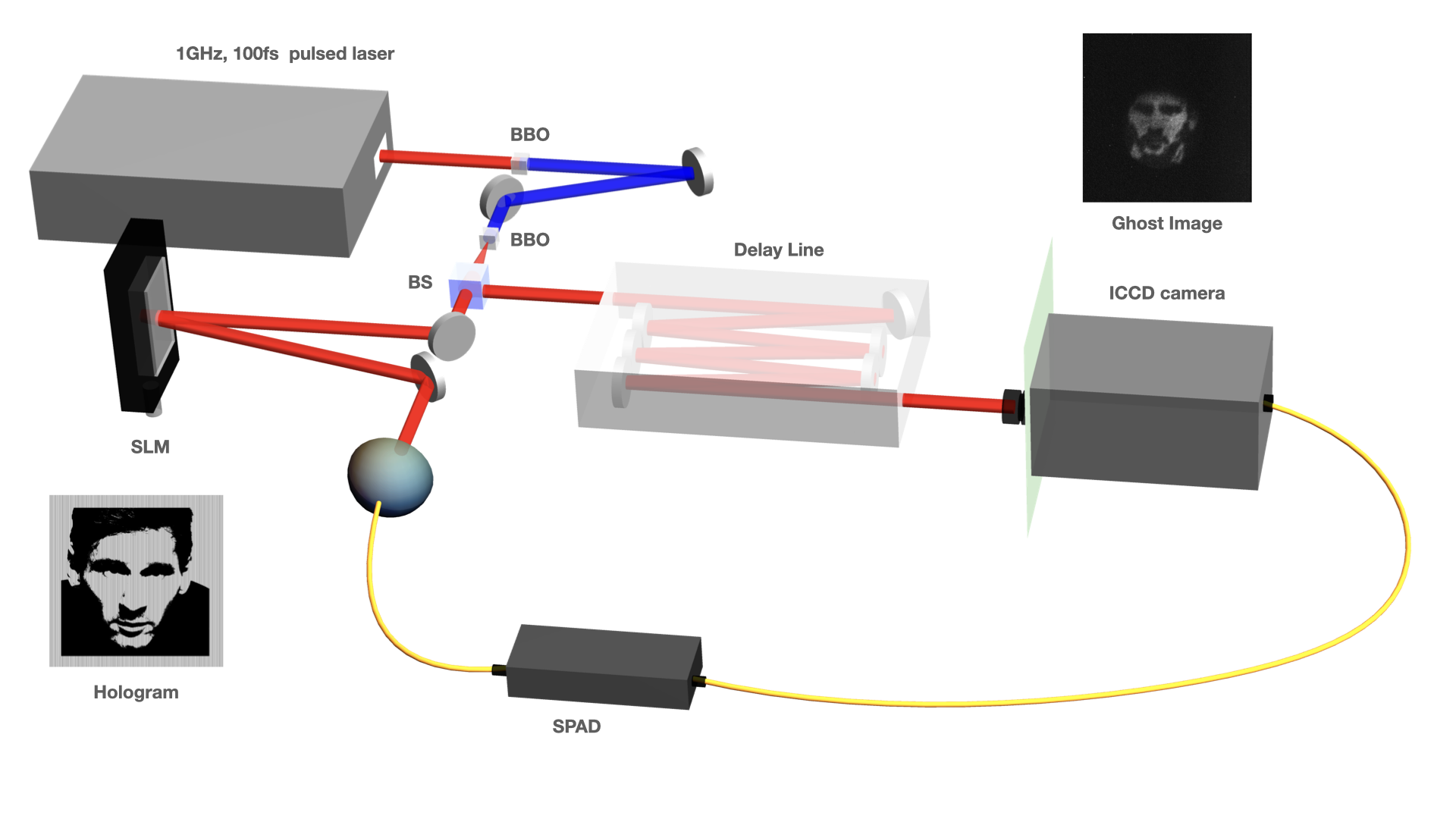}
    \caption{Simplified schematic of the experimental setup for Quantum Ghost Imaging. A 1 GHz, 100 fs laser is used to pump a nonlinear crystal (BBO) for second harmonic generation. The second harmonic beam is used to pump a Type-I bismuth triborate (BiBO) crystal for entangled photon pair generation. One of the photons is sent to a Spatial Light Modulator, a liquid crystal device, on which images of human faces are superposed with a diffraction grating. The second photon is sent to a camera through an image preserving delay line where the image of the object is formed. Figure legends: BiBO - 0.5-mm-thick bismuth triborate crystal; BS - Beamsplitter; BS - Beam splitter; SPAD - Single Photon Avalanche Diode; ICCD - Intensified CCD camera.}
    \label{fig:s1}
\end{figure*}

We elaborate on the experimental details of the image acquisition for the quantum pattern recognition protocol. Spatially correlated photon pairs, usually called signal and idler photons, are generated from a Spontaneous Parametric Down Conversion Process (SPDC) by pumping a nonlinear crystal. Utilizing the position and momentum correlations in these down converted photon pairs, one can non-locally obtain an image of an object that interacted only with the idler photons. The experimental setup we use is similar to a conventional ghost-imaging setup, see Fig.~\ref{fig:s1}, with our object being a hologram placed in a Spatial Light Modulator (SLM), a liquid crystal device. We use a 1~GHZ, 100~fs pulsed laser to pump a nonlinear crystal, $\beta$-Barium Borate (BBO), for generating a second harmonic output. We then use the second harmonic beam to pump a Type-I bismuth triborate (BiBO) crystal for the down conversion of photon pairs. The generated signal and idler pairs are split into two paths, i.e. the object arm (idler) and the camera arm (signal), via a 50:50 Beamsplitter (BS). The idler photon interacts with the SLM, on which we display the holograms created by superimposing the original face image with a diffraction grating. The grating sends only the desired photons from the incident beam to the the first order, which then are coupled to a Single Mode Fibre (SMF) and sent into a Single Photon Avalanche Diode (SPAD) detector which can be used to trigger the collection of the photons in the Intensified CCD (ICCD) camera. The images obtained that are shown in Fig.~\ref{Fig:fig2} were taken with 0.5 s exposure accumulated over 300 frames.

\newpage
\section{Quantum Computation of Trace}

Here, we suggest an adder algorithm~\cite{ruiz2017quantum} to compute the trace of a matrix via adding the diagonal elements of the matrix A. This operator is mainly based on quantum Fourier transform (QFT) and inverse QFT (i.e. QFT$^{-1}$). The algorithm should process the binary forms of the diagonal. For example, the binary representation of the diagonal elements $a_{11}$ and $a_{22}$ of matrix A are respectively $a_{11}=\alpha_12^{n-1}+\alpha_22^{n-2}+\ldots+\alpha_n2^0$ and $a_{22}=\beta_12^{n-1}+\beta_22^{n-2}+\ldots+\beta_n2^0$, which are $\vert a_{11} \rangle = \vert \alpha_1\rangle \otimes \vert \alpha_2\rangle \otimes \ldots \vert \alpha_n\rangle$ and $\vert a_{22} \rangle = \vert \beta_1\rangle \otimes \vert \beta_2\rangle \otimes \ldots\vert \beta_n\rangle $ in the form of quantum kets. The QFT operation on binary state is $\text{QFT}\vert a \rangle = \frac{1}{\sqrt N}\sum_{k=0}^{N-1}e^{\frac{i2\pi a k}{N}}\vert k \rangle$ and the operation of QFT$^{-1}$ is
$\text{QFT}^{-1}\vert k \rangle = \frac{1}{\sqrt N}\sum_{a =0}^{N-1}e^{\frac{-i2\pi a k}{N}}\vert a \rangle$\cite{ruiz2017quantum}.
For simplicity, we introduce a representation for QFT as $$\vert \Phi (a) \rangle = \text{QFT}\vert a \rangle = \frac{1}{\sqrt N}\sum_{k=0}^{N-1}e^{\frac{i2\pi a k}{N}}\vert k \rangle,$$ so, we can write $$\text{QFT}^{-1}\vert \Phi (a) \rangle = \text{QFT}^{-1}\text{QFT}\vert a \rangle = \vert a \rangle.$$

In order to calculate the trace, we need to add all diagonal elements $\vert a_{11} \rangle+\vert a_{22} \rangle+\ldots+\vert a_{NN} \rangle$ to have the ket include the value of trace as $\vert a_{11} + a_{22} +\ldots+ a_{NN} \rangle$. We introduce the operator $\Sigma$ that adds two elements $a_{11}$ and $a_{22}$ in the form of $\vert a_{11}+a_{22} \rangle$ as follows:
$$\Sigma (\vert a_{11} \rangle \vert \Phi (a_{22}) \rangle)= \vert \Phi (a_{11}+a_{22}) \rangle.$$
Then, after the operation of QFT$^{-1}$ we obtain
$$\text{QFT}^{-1}(\vert \Phi (a_{11}+a_{22}) \rangle)=\vert a_{11}+a_{22} \rangle.$$

\begin{figure*}
\centering
\includegraphics[width=0.7\linewidth]{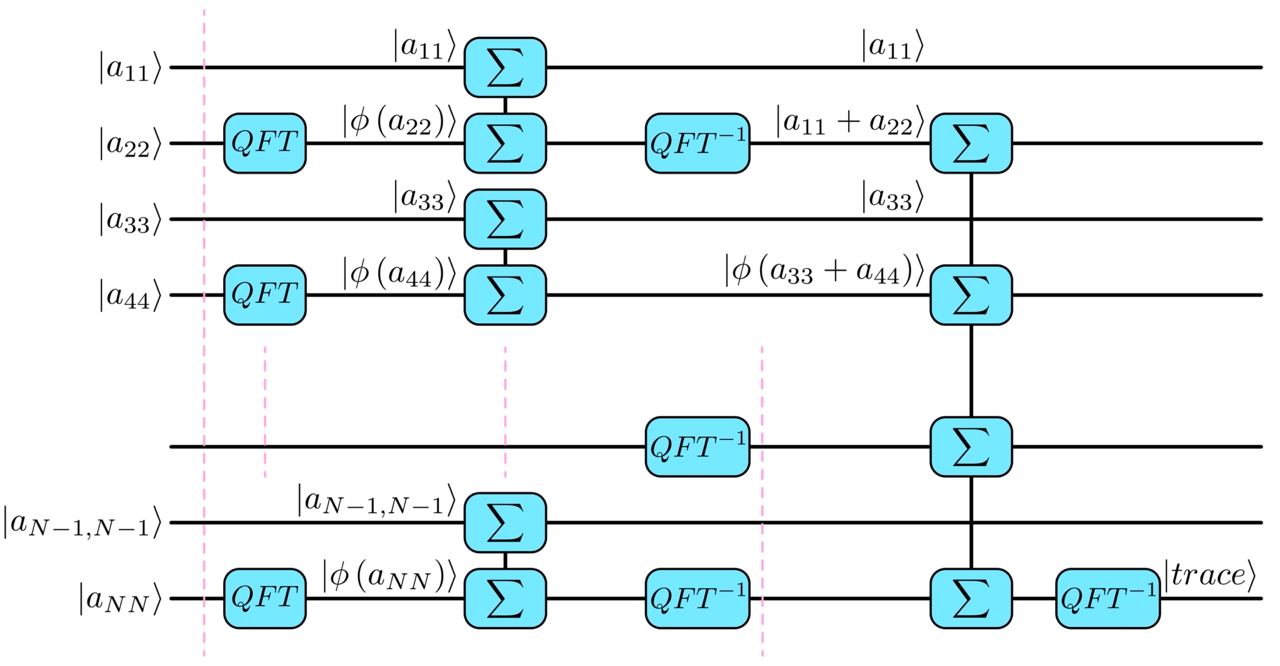}
\caption{Quantum circuit for the trace calculation of sparse matrix.}
\label{fig:s2}
\end{figure*}

By continuation of this method for the all diagonal elements, the trace can be obtained. The quantum protocol for computation of trace is depicted in Fig.~\ref{fig:s2}, in which the input is $\vert a_{11} \rangle \vert a_{22}\rangle\ldots\vert a_{NN} \rangle$ and the output is $\vert \Phi (a_{11}+a_{22}+\ldots+a_{NN}) \rangle=\vert \Phi (Tr(A)) \rangle$. Finally, by an operation of QFT$^{-1}$ we can get $\vert Tr(A) \rangle$. The whole process based on QFT and QFT$^{-1}$ has a complexity $\log{N}$.

\newpage
\section{Quantum computation of sparse matrix determinants}

Our algorithm for computation of determinant is clarified in the following subsections as inputs, algorithm boxes, and algorithm steps: 
\subsubsection{Inputs}
\begin{itemize}
\item  Sparse matrix $A$
\item $ \vert 0\rangle^{\otimes n} \vert \Psi \rangle $ as the input in QPE
\item $  \vert 0 \rangle$ as the ancilla for rotation operator
\item $  \vert 0\rangle ^{\otimes N}$ as the memory register for multiplication operator
\end{itemize}
\subsubsection{Algorithm Boxes}
\begin{itemize}
\item {$\text{QPE}$ is the quantum phase estimation subroutine composed of $H^{\otimes n}$, $\text{CU}$ (i.e. controlled-U) and inverse quantum Fourier transform ($\text{QFT}^{-1}$)}
\item {Rotation operation ($\mathbb{R}$})
\item {$\text{(QPE)}^{-1}$ is the inverse operation of $\text{QPE}$, composed of $H^{\otimes n}$, $\text{CU}^{\dagger}$ and quantum Fourier transform ($\text{QFT}$)}
\item {$\Pi$ is a multiplication operation}
\end{itemize}

\begin{figure*}
\centering
\includegraphics[width=.7\linewidth]{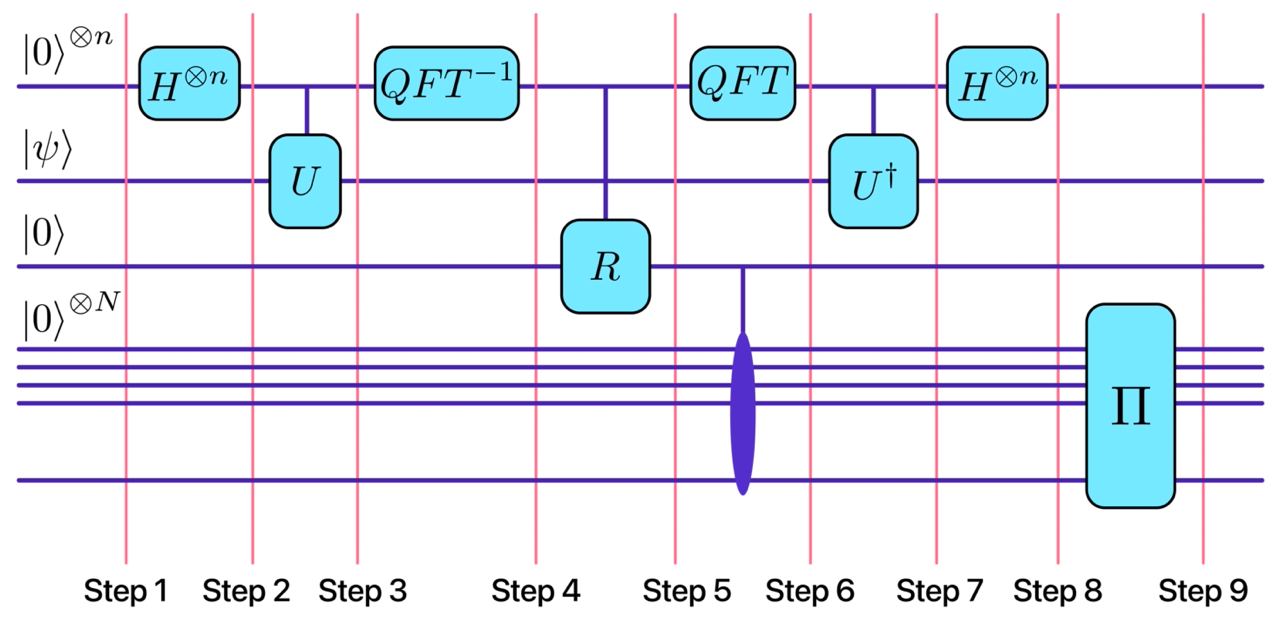}
\caption{Quantum circuit of determinant calculation of sparse matrix}
\label{fig:figs3}
\end{figure*}

The matrix $A$ can be exponentiated as the unitary operator $U = e^{\frac{2\pi iA}{2^n}}$ with logarithmic complexity~\cite{lloyd2014quantum} in which $n$ is the precision. This unitary operator is used in the controlled-U (i.e. CU) part of QPE, and $\vert \Psi \rangle$ is the superposition of the eigenvectors of $A$ in the form of $ \vert \Psi \rangle = \sum \beta_j  \vert u _ j \rangle$. Figure~\ref{fig:figs3} is the representation of the quantum protocol for the computation of matrix determinants. The following steps are based on the steps shown in Fig.~\ref{fig:figs3}.\\

\noindent STEP 1:\\
The initial state of the algorithm is:
\begin{equation}
\vert 0 \rangle^{\otimes n }\vert \Psi \rangle \vert 0 \rangle \vert 0 \rangle^{\otimes N}.
\end{equation}

\noindent STEP 2:\\
After the operation of $n$ Hadamard gates, i.e. $H^{\otimes n}$, we have:
\begin{equation}
\frac{1}{2^{\frac{n}{2}}}\sum_{y_1,\ldots,y_n=0}^{1} \vert y_1\ldots y_n \rangle \vert \Psi \rangle \vert 0 \rangle \vert 0 \rangle^{\otimes N}.
\end{equation}

\noindent STEP 3:\\
In this step, let us apply the controlled-U (CU) operation:
\begin{equation}
\frac{1}{2^{\frac{n}{2}}}\sum_{y=0}^{2^n-1} \sum_{j=1}^{N} \beta_j e^{\frac{2\pi i \lambda_jy}{2^n}}\vert y \rangle \vert u_j \rangle
\vert 0 \rangle \vert 0 \rangle^{\otimes N},
\end{equation}
where $y=\sum_{l=1}^{n}y_l2^{n-l}$ and $\lambda_j$'s are the eigenvalues of matrix $A$.\\

\noindent STEP 4:\\
Here, we apply the inverse Fourier Transform,
\begin{equation}
\frac{1}{2^n}\sum_{y=0}^{2^n-1}\sum_{k=0}^{2^n-1} \sum_{j=1}^{N} \beta_j e^{2\pi i \left(\frac{ \lambda_j-k}{2^n}\right)y}\vert k \rangle \vert u_j \rangle
\vert 0 \rangle \vert 0 \rangle^{\otimes N}.
\end{equation}
For a single $k$ among the all possible values, we have $ \lambda_j -k=0$, where $ \lambda_j=x=k$. The other terms will be set to zero. Thus, the notation $k$ is changed to the notation $x$:
\begin{equation}
 \frac{1}{2^n}\sum_{y=0}^{2^n-1} e^{2\pi i \left(\frac{ \lambda_j-x}{2^n}\right)y}=1
\end{equation}
In this case, the state becomes:
$$ \sum_{x=0}^{2^n-1} \sum_{j=1}^{N} \beta_j \vert \lambda_j \rangle \vert u_j \rangle \vert 0 \rangle \vert 0 \rangle^{\otimes N }= \sum_{x_1 \ldots x_n=0}^{1} \sum_{j=1}^{N} \beta_j \vert x_1x_2\ldots x_n \rangle \vert u_j \rangle \vert 0 \rangle \vert 0 \rangle^{\otimes N},$$
where $$ \lambda_j=\sum_{l=0}^{n}x_l2^{n-l}=2^n\sum_{l=0}^{n}x_l2^{-l}=2^n\tilde \lambda_j\,.$$\\

\noindent STEP 5:\\
In this step, we apply the rotation operation $\mathbb{R}_l=e^{\frac{i\mathbb{\sigma}_y}{2^l}}$, where $\sigma_y$ is the Pauli matrix y, which acts on the output of QFT$^{-1}$:
\begin{equation}
 \sum_{x_1\ldots x_n=0}^{1} \sum_{j=1}^{N} \beta_j \vert x_1\ldots x_n \rangle \vert u_j \rangle (\sqrt{1- \tilde \lambda^2_j} \vert 0 \rangle +\tilde \lambda_j \vert 1 \rangle).
\end{equation}

\noindent STEP 6:\\
Applying the quantum Fourier transform (QFT) results in:
\begin{align}
\frac{1}{2^{\frac{n}{2}}}\sum_{y=0}^{2^n-1} \sum_{j=1}^{N} \beta_j e^{\frac{2\pi i \lambda_j y}{2^n}}\vert y \rangle  \vert u_j \rangle (\sqrt{1- \tilde \lambda^2_j} \vert 0 \rangle +\tilde \lambda_j  \vert 1 \rangle).
\end{align}

\noindent STEP 7: \\
In this step, we apply the $\text{CU}^\dagger$ operator,
\begin{equation}
\frac{1}{2^{\frac{n}{2}}}\sum_{y=0}^{2^n-1} \sum_{j=1}^{N} \beta_j \vert y \rangle \vert u_j \rangle
(\sqrt{1- \tilde \lambda^2_j} \vert 0 \rangle +\tilde \lambda_j  \vert 1 \rangle).
\end{equation}

\noindent STEP 8:\\
As $\vert u_j \rangle$'s are known, we repeat the algorithm N times, each time for a specific $\vert u_j \rangle$, and consequently we obtain the following state:
\begin{equation}
\vert 0 \rangle^{\otimes n }\vert \Psi \rangle \prod _{j=1}^{n}(\sqrt{1- \tilde \lambda^2_j} \vert 0 \rangle +\tilde \lambda_j \vert 1 \rangle).
\end{equation}

Now, the goal is to measure the multiplication of $\tilde\lambda_j$'s in the output of multiplication operation;
\begin{eqnarray}
\prod _j ^ N & & \left(\sqrt{1- \tilde \lambda_j ^2} \vert 0 \rangle +  \tilde \lambda_j \vert 1 \rangle \right)=
\left(  \sqrt{1- \tilde \lambda_1 ^2} \vert 0 \rangle +  \tilde \lambda_1 \vert 1 \rangle \right)\ldots
 \left(  \sqrt{1- \tilde \lambda_N ^2} \vert 0 \rangle +  \tilde \lambda_N \vert 1 \rangle \right)\\
&=&\left(  \sqrt{1- \tilde \lambda_1 ^2} \right) \ldots \left(  \sqrt{1- \tilde \lambda_N ^2} \right) \vert 00 \ldots 0 \rangle +
\left(  \sqrt{1- \tilde \lambda_1 ^2} \right)(\tilde \lambda_2)\ldots \vert 0100\ldots0\rangle + \ldots
 + \overbrace{(\tilde \lambda_1)(\tilde \lambda_2)\ldots(\tilde \lambda_N)} ^ { \tilde \lambda_1 ... \tilde \lambda_N}\overbrace{\vert 111\ldots1 \rangle}^{N}.\quad\qquad
\end{eqnarray}
The coefficient of the state $\vert \underbrace{1 1 \ldots1}_N \rangle $ is the term $\tilde \lambda_1\tilde \lambda_2 \ldots\tilde \lambda_N$ in the output of the multiplier, which can be obtained via a weak measurement without collapsing other lines. As $\lambda_j=2^n\tilde \lambda_j$, we can obtain the determinant of $A$ (i.e. $\lambda_1\lambda_2\ldots\lambda_N$) via relation ${(2^n)}^N\tilde \lambda_1\tilde \lambda_2 \ldots\tilde \lambda_N=\lambda_1\lambda_2\ldots\lambda_N=det(A)$.

\subsection{Proof of the steps}
The proof for each of the eight steps, described above, are given in details in the following expressions.\newline

\noindent STEP 2:
\begin{eqnarray}
|\Psi_2 \rangle &=& (H^{\otimes n}_2\otimes \mathbb{I}_N\otimes\mathbb{I}_2 \otimes \mathbb{I}^{\otimes N}_2)\vert 0 \rangle^{\otimes n }\vert \Psi \rangle \vert 0 \rangle \vert 0 \rangle^{\otimes N}=\frac{1}{2^{\frac{n}{2}}}\sum_{y=0}^{2^n-1} \vert y \rangle \vert \Psi \rangle \vert 0 \rangle \vert 0 \rangle^{\otimes N }\nonumber\\
&=&\frac{1}{2^{\frac{n}{2}}}\sum_{y_1...y_n=0}^{1} \vert y_1...y_n \rangle \vert \Psi \rangle \vert 0 \rangle \vert 0 \rangle^{\otimes N }
\end{eqnarray}

\noindent STEP 3:
\begin{eqnarray}
|\Psi_3\rangle &=& \prod _{l=1}^{n}(\mathbb{I}^{\otimes l-1}_2\otimes \vert 0 \rangle \langle 0 \vert \otimes \mathbb{I}^{\otimes n-l}_2 \otimes \mathbb{I}_N\otimes\mathbb{I}_2 \otimes \mathbb{I}^{\otimes N}_2+
\mathbb{I}^{\otimes l-1}_2\otimes \vert 1 \rangle \langle 1 \vert \otimes \mathbb{I}^{\otimes n-l}_2 \otimes U^{2^{n-l}}\otimes\mathbb{I}_2\otimes \mathbb{I}^{\otimes N}_2)  
\frac{1}{2^{\frac{n}{2}}} \times \nonumber \\ 
&\times& \sum_{y_1...y_n=0}^{1} \vert y_1...y_n \rangle \sum_{j=1}^{N} \beta_j \vert u_j \rangle \vert 0 \rangle \vert 0 \rangle^{\otimes N} \nonumber \\
&=&\frac{1}{2^{\frac{n}{2}}}\sum_{y_1...y_n=0}^{1} \sum_{j=1}^{N} \beta_j \prod _{l=1}^{n} (\text{CU})_l\vert y_1...y_n \rangle \vert u_j \rangle \vert 0 \rangle \vert 0 \rangle^{\otimes N }\nonumber\\
&=& \frac{1}{2^{\frac{n}{2}}}( \sum_{y_1...y_n=0}^{1} \sum_{j=1}^{N} \beta_j \prod _{l=1}^{n} (\delta_{0,y_l}+\delta_{1,y_l}e^{2\pi i \tilde \lambda_j2^{n-l}})\vert y_1...y_n \rangle \vert u_j \rangle
\vert 0 \rangle) \vert 0 \rangle^{\otimes N }\nonumber\\
&=& \frac{1}{2^{\frac{n}{2}}}( \sum_{y_1...y_n=0}^{1} \sum_{j=1}^{N} \beta_j \prod _{l=1}^{n} e^{2\pi i \tilde \lambda_jy_l2^{n-l}}\vert y_1...y_n \rangle \vert u_j \rangle\vert 0 \rangle)\vert 0 \rangle^{\otimes N}\nonumber\\
&=& \frac{1}{2^{\frac{n}{2}}}( \sum_{y_1...y_n=0}^{1} \sum_{j=1}^{N} \beta_j e^{2\pi i \tilde \lambda_j\sum_{l=1}^{n}y_l2^{n-l}}\vert y_1...y_n \rangle \vert u_j \rangle\vert 0 \rangle)\vert 0 \rangle^{\otimes N}\nonumber\\
&=&\frac{1}{2^{\frac{n}{2}}}\sum_{y=0}^{2^n-1} \sum_{j=1}^{N} \beta_j e^{2\pi i \tilde \lambda_jy}\vert y \rangle \vert u_j \rangle \vert 0 \rangle \vert 0 \rangle^{\otimes N}
\end{eqnarray}

\noindent STEP 4:
\begin{eqnarray}
|\Psi_4\rangle &=& (\text{QFT}^{-1}\otimes \mathbb{I}_N\otimes\mathbb{I}_2 \otimes \mathbb{I}^{\otimes N}_2)\frac{1}{2^{\frac{n}{2}}}\sum_{y=0}^{2^n-1} \sum_{j=1}^{N} \beta_j e^{2\pi i \tilde \lambda_jy}\vert y \rangle \vert u_j \rangle \vert 0 \rangle \vert 0 \rangle^{\otimes N } \nonumber\\
&=& \frac{1}{2^{\frac{n}{2}}}\sum_{y=0}^{2^n-1} \sum_{j=1}^{N} \beta_j e^{2\pi i \tilde \lambda_jy}(\text{QFT}^{-1}\vert y \rangle) \vert u_j \rangle\vert 0 \rangle \vert 0 \rangle^{\otimes N} \nonumber\\
&=& \frac{1}{2^{\frac{n}{2}}}\sum_{y=0}^{2^n-1} \sum_{j=1}^{N} \beta_j e^{2\pi i \tilde \lambda_jy}(\frac{1}{2^{\frac{n}{2}}}\sum_{k=0}^{2^n-1}e^{-2\pi i\frac{k}{2^n} y}\vert k \rangle \langle y \vert y \rangle) \vert u_j \rangle \vert 0 \rangle \vert 0 \rangle^{\otimes N} \nonumber\\
&=&\frac{1}{2^n}\sum_{y=0}^{2^n-1}\sum_{k=0}^{2^n-1} \sum_{j=1}^{N} \beta_j e^{2\pi i (\tilde \lambda_j-\frac{k}{2^n})y}\vert k \rangle \vert u_j \rangle\vert 0 \rangle \vert 0 \rangle^{\otimes N} \nonumber\\
&=&\frac{1}{2^n}\sum_{y=0}^{2^n-1}\sum_{k=0}^{2^n-1} \sum_{j=1}^{N} \beta_j e^{2\pi i (\frac{ \lambda_j-k}{2^n})y}\vert k \rangle \vert u_j \rangle \vert 0 \rangle \vert 0 \rangle^{\otimes N }
\end{eqnarray}

\begin{equation}
|\Psi_4\rangle = \sum_{x=0}^{2^n-1} \sum_{j=1}^{N} \beta_j \vert \lambda_j \rangle \vert u_j \rangle \vert 0 \rangle \vert 0 \rangle^{\otimes N }= \sum_{x_1...x_n=0}^{1} \sum_{j=1}^{N} \beta_j \vert x_1x_2...x_n \rangle \vert u_j \rangle \vert 0 \rangle \vert 0 \rangle^{\otimes N }
\end{equation}

\noindent STEP 5:
\begin{eqnarray}
|\Psi_5\rangle &=& \prod _{l=1}^{n}(\mathbb{I}^{\otimes n-l+1}_2\otimes \vert 0 \rangle \langle 0 \vert \otimes \mathbb{I}^{\otimes l}_2 \otimes \mathbb{I}_N\otimes\mathbb{I}_2 \otimes \mathbb{I}^{\otimes N}_2 +\mathbb{I}^{\otimes n-l+1}_2\otimes \vert 1 \rangle \langle 1 \vert \otimes \mathbb{I}^{\otimes l}_2 \otimes \mathbb{I}_N\otimes \mathbb{R}_{n-l})\otimes \mathbb{I}^{\otimes N}_2  \times \nonumber \\
&\times&\sum_{x_1...x_n=0}^{1} \vert x_1...x_n \rangle \sum_{j=1}^{N} \beta_j \vert u_j \rangle \vert 0 \rangle \nonumber\\
&=&\sum_{x_1...x_n=0}^{1} \sum_{j=1}^{N} \beta_j \prod _{l=1}^{n} (\text{CR})_{n-l+1}\vert x_1...x_n \rangle \vert u_j \rangle \vert 0 \rangle \vert 0 \rangle^{\otimes N } \nonumber\\
&=&  \sum_{x_1...x_n=0}^{1} \sum_{j=1}^{N} \beta_j \vert x_1...x_n \rangle \vert u_j \rangle(\prod _{l=1}^{n} (\delta_{0,x_{n-l+1}}+\delta_{1,x_{n-l+1}}e^{\frac{i\mathbf{\sigma}_y}{2^{n-l+1}}}) \vert 0 \rangle)\vert 0 \rangle^{\otimes N }\nonumber\\
&=&  \sum_{x_1...x_n=0}^{1} \sum_{j=1}^{N} \beta_j \vert x_1...x_n \rangle \vert u_j \rangle (\prod _{l=1}^{n} e^{i\mathbf{\sigma}_y x_{n-l+1}2^{-(n-l+1)}} \vert 0 \rangle)\vert 0 \rangle^{\otimes N } \nonumber\\
&=&\sum_{x_1...x_n=0}^{1} \sum_{j=1}^{N} \beta_j \vert x_1...x_n \rangle \vert u_j \rangle (e^{ i\mathbf{\sigma}_y \sum_{l=1}^{n}x_{n-l+1}2^{-(n-l+1)}}\vert 0 \rangle)\vert 0 \rangle^{\otimes N }\nonumber\\
&=& \sum_{x_1...x_n=0}^{1} \sum_{j=1}^{N} \beta_j \vert x_1...x_n \rangle \vert u_j \rangle (e^{ i \mathbf{\sigma}_y\tilde \lambda_j}\vert 0 \rangle) \vert 0 \rangle^{\otimes N}.
\end{eqnarray}

\noindent STEP 6:
\begin{eqnarray}
|\Psi_6\rangle &=& (\text{QFT}\otimes \mathbb{I}_N\otimes\mathbb{I}_2\otimes \mathbb{I}^{\otimes N}_2)\sum_{x_1...x_n=0}^{1} \sum_{j=1}^{N} \beta_j \vert x_1...x_n \rangle \vert u_j \rangle (\sqrt{1- \tilde \lambda^2_j} \vert 0 \rangle +\tilde \lambda_j \vert 1 \rangle)\nonumber \\
&=& (\text{QFT} \sum_{x_1...x_n=0}^{1} \vert x_1...x_n \rangle) \sum_{j=1}^{N}\beta_j \vert u_j \rangle (\sqrt{1- \tilde \lambda^2_j} \vert 0 \rangle +\tilde \lambda_j \vert 1 \rangle) \nonumber\\
&=& (\frac{1}{2^{\frac{n}{2}}}\sum_{y=0}^{2^n-1}e^{2\pi i\frac{k}{2^n} y}\vert y \rangle \langle k \vert \sum_{x_1...x_n=0}^{1} \vert x_1...x_n \rangle) \sum_{j=1}^{N} \beta_j \vert u_j \rangle (\sqrt{1- \tilde \lambda^2_j} \vert 0 \rangle +\tilde \lambda_j  \vert 1 \rangle)\nonumber\\
&=&(\frac{1}{2^{\frac{n}{2}}}\sum_{y=0}^{2^n-1}e^{2\pi i\frac{k}{2^n} y}\vert y \rangle \sum_{x_1...x_n=0}^{1} \langle k \vert x_1...x_n \rangle) \sum_{j=1}^{N} \beta_j \vert u_j \rangle (\sqrt{1- \tilde \lambda^2_j} \vert 0 \rangle +\tilde \lambda_j  \vert 1 \rangle)\nonumber\\
&=&(\frac{1}{2^{\frac{n}{2}}}\sum_{y=0}^{2^n-1}e^{2\pi i\frac{k}{2^n} y}\vert y \rangle \sum_{x_1...x_n=0}^{1} \delta_{k, x_1...x_n}) \sum_{j=1}^{N} \beta_j \vert u_j \rangle (\sqrt{1- \tilde \lambda^2_j} \vert 0 \rangle +\tilde \lambda_j  \vert 1 \rangle)\nonumber\\
&=& (\frac{1}{2^{\frac{n}{2}}}\sum_{y=0}^{2^n-1}e^{2\pi i\frac{k}{2^n} y}\vert y \rangle \delta_{k, \lambda_j}) \sum_{j=1}^{N} \beta_j \vert u_j \rangle (\sqrt{1- \tilde \lambda^2_j} \vert 0 \rangle +\tilde \lambda_j  \vert 1 \rangle)\nonumber \\
&=&\frac{1}{2^{\frac{n}{2}}}\sum_{y=0}^{2^n-1}e^{2\pi i\frac{ \lambda_j}{2^n} y}\vert y \rangle \sum_{j=1}^{N} \beta_j \vert u_j \rangle (\sqrt{1- \tilde \lambda^2_j} \vert 0 \rangle +\tilde \lambda_j  \vert 1 \rangle)\nonumber\\
&=&\frac{1}{2^{\frac{n}{2}}}\sum_{y=0}^{2^n-1} \sum_{j=1}^{N} \beta_j e^{2\pi i\frac{\lambda_j}{2^n} y}\vert y \rangle  \vert u_j \rangle (\sqrt{1- \tilde \lambda^2_j} \vert 0 \rangle +\tilde \lambda_j  \vert 1 \rangle)
\end{eqnarray}

\noindent STEP 7:
\begin{eqnarray}
|\Psi_7\rangle &=& \prod _{l=1}^{n}(\mathbb{I}^{\otimes l-1}_2\otimes \vert 0 \rangle \langle 0 \vert \otimes \mathbb{I}^{\otimes n-l}_2 \otimes \mathbb{I}_N\otimes\mathbb{I}_2 \otimes \mathbb{I}^{\otimes N}_2+\mathbb{I}^{\otimes l-1}_2\otimes \vert 1 \rangle \langle 1 \vert \otimes \mathbb{I}^{\otimes n-l}_2 \otimes (U^\dagger)^{2^{n-l}}\otimes\mathbb{I}_2 \otimes \mathbb{I}^{\otimes N}_2) \times \nonumber\\ 
&\times& \frac{1}{2^{\frac{n}{2}}}\sum_{y=0}^{2^n-1} \sum_{j=1}^{N} \beta_j e^{2\pi i\frac{ \lambda_j}{2^n} y}\vert y \rangle  \vert u_j \rangle (\sqrt{1- \tilde \lambda^2_j} \vert 0 \rangle +\tilde \lambda_j  \vert 1 \rangle)\nonumber\\
&=&\frac{1}{2^{\frac{n}{2}}}\sum_{y_1...y_n=0}^{1} \sum_{j=1}^{N} \beta_j e^{2\pi i\frac{ \lambda_j}{2^n} y} \prod _{l=1}^{n} (\text{CU}^\dagger)_l\vert y_1...y_n \rangle \vert u_j \rangle(\sqrt{1- \tilde \lambda^2_j} \vert 0 \rangle +\tilde \lambda_j  \vert 1 \rangle)\nonumber\\
&=& \frac{1}{2^{\frac{n}{2}}}( \sum_{y_1...y_n=0}^{1} \sum_{j=1}^{N} \beta_j e^{2\pi i\frac{ \lambda_j}{2^n} y} \prod _{l=1}^{n} (\delta_{0,y_l}+\delta_{1,y_l}e^{-2\pi i \lambda_j2^{n-l}})\vert y_1...y_n \rangle \vert u_j \rangle (\sqrt{1- \tilde \lambda^2_j} \vert 0 \rangle +\tilde \lambda_j  \vert 1 \rangle)\nonumber\\
&=& \frac{1}{2^{\frac{n}{2}}}( \sum_{y_1...y_n=0}^{1} \sum_{j=1}^{N} \beta_j e^{2\pi i\frac{ \lambda_j}{2^n} y} \prod _{l=1}^{n} e^{-2\pi i \lambda_jy_l2^{n-l}}\vert y_1...y_n \rangle \vert u_j \rangle (\sqrt{1- \tilde \lambda^2_j} \vert 0 \rangle +\tilde \lambda_j  \vert 1 \rangle)\nonumber\\
&=& \frac{1}{2^{\frac{n}{2}}}( \sum_{y_1...y_n=0}^{1} \sum_{j=1}^{N} \beta_j e^{2\pi i\frac{ \lambda_j}{2^n} y} e^{-2\pi i \lambda_j\sum_{l=1}^{n}y_l2^{n-l}}\vert y_1...y_n \rangle \vert u_j \rangle (\sqrt{1- \tilde \lambda^2_j} \vert 0 \rangle +\tilde \lambda_j  \vert 1 \rangle)\nonumber\\
&=& \frac{1}{2^{\frac{n}{2}}}( \sum_{y_1...y_n=0}^{1} \sum_{j=1}^{N} \beta_j e^{2\pi i\frac{ \lambda_j}{2^n} y} e^{-2\pi i \tilde \lambda_jy}\vert y_1...y_n \rangle \vert u_j \rangle (\sqrt{1- \tilde \lambda^2_j} \vert 0 \rangle +\tilde \lambda_j  \vert 1 \rangle)\nonumber\\
&=& \frac{1}{2^{\frac{n}{2}}}\sum_{y=0}^{2^n-1} \sum_{j=1}^{N} \beta_j e^{\frac{2\pi i}{2^n} ( \lambda_j- \lambda_j)y}\vert y \rangle \vert u_j \rangle(\sqrt{1- \tilde \lambda^2_j} \vert 0 \rangle +\tilde \lambda_j  \vert 1 \rangle)\nonumber\\
&=& \frac{1}{2^{\frac{n}{2}}}\sum_{y=0}^{2^n-1} \sum_{j=1}^{N} \beta_j \vert y \rangle \vert u_j \rangle (\sqrt{1- \tilde \lambda^2_j} \vert 0 \rangle +\tilde \lambda_j  \vert 1 \rangle).
\end{eqnarray}

\noindent STEP 8:
\begin{eqnarray}\label{eq8}
|\Psi_8\rangle &=& (H^{\otimes n}_2\otimes \mathbb{I}_N\otimes\mathbb{I}_2 \otimes \mathbb{I}^{\otimes N}_2)\frac{1}{2^{\frac{n}{2}}}\sum_{y=0}^{2^n-1} \sum_{j=1}^{N} \beta_j \vert y \rangle \vert u_j \rangle (\sqrt{1- \tilde \lambda^2_j} \vert 0 \rangle +\tilde \lambda_j  \vert 1 \rangle) \nonumber\\
&=& |0\rangle^{\otimes n} \sum_{j=1}^{N} \beta_j \vert u_j \rangle (\sqrt{1- \tilde \lambda^2_j} \vert 0 \rangle +\tilde \lambda_j  \vert 1 \rangle)
\end{eqnarray}

\begin{eqnarray}\nonumber
\prod _j ^ N & & \left(  \sqrt{1- \tilde \lambda_j ^2} \vert 0 \rangle +  \tilde \lambda_j \vert 1 \rangle \right)=
\left(  \sqrt{1- \tilde \lambda_1 ^2} \vert 0 \rangle + \tilde \lambda_1 \vert 1 \rangle \right)  \ldots
 \left(  \sqrt{1- \tilde \lambda_N ^2} \vert 0 \rangle +  \tilde \lambda_N \vert 1 \rangle \right) \\
&=&\left(  \sqrt{1- \tilde \lambda_1 ^2} \right) \ldots \left(  \sqrt{1- \tilde \lambda_N ^2} \right) \vert 00 \ldots 0 \rangle +
\left(  \sqrt{1- \tilde \lambda_1 ^2} \right)(\tilde \lambda_2)\ldots \vert 0100\ldots0\rangle + \ldots
 + \overbrace{(\tilde \lambda_1)(\tilde \lambda_2)\ldots(\tilde \lambda_N)} ^ { \tilde \lambda_1 \ldots \tilde \lambda_N}\overbrace{\vert 111\ldots1 \rangle}^{N}\quad\qquad
\end{eqnarray}

\end{document}